\documentclass[fleqn,usenatbib]{mnras}

\usepackage{newtxtext,newtxmath}
\DeclareRobustCommand{\VAN}[3]{#2}
\let\VANthebibliography\thebibliography%
\def\thebibliography{\DeclareRobustCommand{\VAN}[3]{##3}\VANthebibliography}

\usepackage{amsmath}
\usepackage{booktabs}
\usepackage[english]{babel}
\usepackage{enumerate}
\usepackage[T1]{fontenc}
\usepackage{graphicx}
\usepackage[utf8]{inputenc}
\usepackage{natbib}
\usepackage[dvipsnames]{xcolor}

\newcommand{\figurewidthtwo}{0.50\linewidth}

\title[X-ray AGN ${M_\mathrm{star}}-{M_\mathrm{halo}}$]%
{The role of scatter and satellites in shaping the large-scale clustering of
X-ray AGN as a function of host galaxy stellar mass}

\author[A\@. Viitanen et al.]{%
A\@. Viitanen,$^{1,2}$\thanks{E-mail: akke.viitanen@helsinki.fi}
V\@. Allevato,$^{3,4,1}$
A\@. Finoguenov,$^{1}$
F\@. Shankar,$^{5}$
and C\@. Marsden$^{5}$
\\
$^{1}${Department of Physics, University of Helsinki, PO Box 64, FI-00014 Helsinki, Finland} \\
$^{2}${Helsinki Institute of Physics, Gustaf Hällströmin katu 2, University of Helsinki, Finland} \\
$^{3}${Scuola Normale Superiore, Piazza dei Cavalieri 7, I-56126 Pisa, Italy} \\
$^{4}${INAF—Osservatorio di Astrofisica e Scienza dello Spazio di Bologna, Bologna, Italy} \\
$^{5}${Department of Physics and Astronomy, University of Southampton, Highfield, SO17 1BJ, UK}
}

\date{%
  Accepted 2021 August 30. %
  Received 2021 August 03; %
  in original form 2021 January 20%
}
\pubyear{2021}

\begin{document}%
\label{firstpage}
\pagerange{\pageref{firstpage}--\pageref{lastpage}}
\maketitle

\begin{abstract}
  The co-evolution between central supermassive black holes (BH), their host
  galaxies, and dark matter halos is still a matter of intense debate. Present
  theoretical models suffer from large uncertainties and degeneracies, for
  example, between the fraction of accreting sources and their characteristic
  accretion rate. In recent work we showed that Active Galactic Nuclei (AGN)
  clustering represents a powerful tool to break degeneracies when analysed in
  terms of mean BH mass, and that AGN bias at fixed stellar mass is largely
  independent of most of the input parameters, such as the AGN duty cycle
  and the mean scaling between BH mass and host galaxy stellar mass. In
  this paper we take advantage of our improved semi-empirical methodology
  and recent clustering data derived from large AGN samples at $z \sim
  1.2$, demonstrate that the AGN bias as a function of host galaxy stellar
  mass is a crucial diagnostic of the BH--galaxy connection, and is highly
  dependent on the scatter around the BH mass--galaxy mass scaling relation
  and on the relative fraction of satellite and central active BHs. Current
  data at $z \sim 1.2$ favour relatively high values of AGN in satellites,
  pointing to a major role of disc instabilities in triggering AGN, unless
  a high minimum host halo mass is assumed. The data are not decisive on
  the magnitude/covariance of the BH-galaxy scatter at $z \sim 1.2$ and
  intermediate host masses $M_\mathrm{star} \lesssim 10^{11}
  \,\mathrm{M}_\odot$. However, future surveys like Euclid/LSST will be pivotal
  in shedding light on the BH--galaxy co-evolution.
\end{abstract}


\section{Introduction}%
\label{sec:introduction}

The presence of supermassive black holes at the centers of virtually every
massive galaxy is an accepted paradigm. The masses of these central black holes
appear to correlate with the properties of their host galaxies
\citep[e.g.][]{ferrarese_ford05,kormendy_ho13,graham15,reines15,shankar19,%
shankar20} with tentative evidence for a link even with their host dark matter
haloes \citep[e.g.][]{ferrarese02,bandara09}. The very existence of such
correlations, which are observed to hold even at higher redshifts
\citep[e.g.][]{shankar09c,cisternas11b,shen15,suh20}, point to a degree of
\emph{co-evolution} between the black holes and their hosts. Thus, unveiling
the shape, dispersion and evolution of these correlations represents a crucial
task in present-day extra-galactic astronomy to acquire a full picture of
galaxy formation and evolution.

Despite decades of observational and theoretical work aimed at deciphering
the nature of the black hole-galaxy scaling relations, the causal link
between the two remains still unsolved. A strong release of energy/momentum
from an accreting central supermassive black hole (BH) shining as an Active
Galactic Nucleus (AGN) naturally predicts a tight and constant correlation
with velocity dispersion \citep[e.g.][]{silk_rees98,granato04}, which has
been recognized as one of the most fundamental property linked to black hole
mass \citep[e.g.][]{bernardi07,shankar19,marsden20}. Merger-driven models
of black hole growth would instead predict a tighter correlation with the
host galaxy stellar mass \citep[e.g.][]{jahnke11,hirschmann10}.

Given the large number of input assumptions in traditional ab initio
cosmological models, a more phenomenological complementary approach
has been introduced in the last decade to constrain the properties
of black holes in a cosmological context. The latter method relies on
identifying the overall populations of active and inactive black holes at
a given redshift on statistical grounds via, e.g.\ continuity equation
techniques \citep[e.g.][]{marconi04,shankar04,shankar13b,aversa15}
and/or semi-empirical mock catalogues tuned to reproduce stellar mass
functions, AGN X-ray luminosity functions and/or AGN clustering properties
\citep[e.g.][]{georgakakis19,aird21,allevato21}

Further constraints on the co-evolution scenario are indeed provided by
studying the spatial clustering of AGN, especially X-ray selected
\citep[e.g.][]{coil09,krumpe10,allevato11,krumpe12,mountrichas_georgakakis12,%
mendez16,powell18,viitanen19,allevato19,powell20}. Clustering
provides an independent way of connecting accreting BHs to their
large-scale environments via the link with their underlying dark matter
halo population. Several clustering studies have argued that AGN
environment could play a significant role in triggering the AGN phase,
highlighting the importance of host dark matter halos in nuclear activity
\citep[e.g.][]{hickox09a,allevato11,fanidakis12,gatti16}.

However useful, interpretation of X-ray AGN clustering results
have suffered from not properly being able to take into account
the host galaxy properties. Indeed what is currently debated is to
which degree AGN clustering can be understood solely in terms of
the underlying host galaxy properties (e.g.\ color, stellar mass
${M_\mathrm{star}}$, star-formation rate), and AGN selection effects
\citep[e.g.][]{hickox09a,mendez16,mountrichas19,viitanen19,powell20}.
Thus of special interest is the host galaxy stellar mass ${M_\mathrm{star}}$
due to its intimate connection with the underlying dark matter halo
\citep[e.g.][]{moster10,leauthaud12,behroozi19}.

In terms of the AGN-hosting dark matter halo connection, despite their
extreme flexibility, all phenomenological AGN models suffer from serious
degeneracies that cannot be easily broken. \citet[][]{shankar20} have shown
that the large-scale clustering of AGN, e.g.\ the bias, as a function of
BH mass represents a powerful framework to constrain the normalization
and slope of the correlation between BH mass and host galaxy stellar mass.
Recently, \citet{allevato21} have found by using AGN mock catalogs built
on observationally derived galaxy - BH scaling relations, AGN duty cycles
(i.e.the probability for a galaxy hosting an active black hole), and Eddington
ratio distributions, that the AGN large-scale bias is a crucial diagnostic
to break degeneracies in the input AGN model. Other groups have, instead,
built the correlation between black hole mass and host halo mass by passing
the scaling relation with host galaxy stellar mass or velocity dispersion
and the AGN duty cycle \citep[e.g.][]{georgakakis19,aird21}. Besides not
allowing for any quantitative assessment of the underlying fundamental
black hole-galaxy scaling relations, limiting the efficacy of these models
in shedding light on the processes controlling the co-evolution of BHs and
their hosts, the latter approach is still not immune to degeneracies between
e.g.\ duty cycles and Eddington ratio distributions.

In particular, \citet{allevato21} have shown at $z \sim 0.1$ that the
clustering at fixed stellar mass is largely independent of the input duty
cycle, Eddington ratio distribution, and the
$M_{\mathrm{star}}$-$M_\mathrm{halo}$ relation. This implies that the
clustering of normal galaxies matches the AGN clustering at a given stellar
mass, provided the AGN hosts are a random subsample of the underlying galaxy
population of the same stellar mass. However, as we will show in this work, the
AGN large-scale bias as a function of stellar mass is mainly set by the scatter
- not so much by the shape - of the BH mass - stellar mass relation and by the
relative fraction of AGN in central and satellite dark matter halos. The role
of the scatter in the BH - host mass relation on the AGN clustering has been
investigated in different studies at high $z \sim 4$
\citep{white08,wyithe09,shankar10,bonoli10}. In particular, \citet{white08}
have shown that the very high bias measured for SDSS quasars at $z \sim 4$ by
\citet{shen07} is in favour of a small scatter in the BH -- galaxy scaling
relation. In fact, the key idea is that any scatter in the input scaling
relation increases the contribution from lower mass and less biased halos, thus
decreasing the observed AGN large-scale bias.

However, these studies focus on high redshifts, where the population of
supermassive BHs that powers the quasars is growing rapidly and inhabits the
rarest, most massive halos. Moreover, they do not investigate the effect
of the scatter on the AGN clustering in different stellar mass bins, also
bypassing additional parameters that might affect the AGN bias normalization.

Instead, in this work we study the concurring effect of the scatter around the
BH mass--galaxy mass relation and of the relative fraction of satellite and
central active BHs on the AGN large-scale clustering, focusing our attention
on the AGN bias as a function of host galaxy stellar mass at intermediate
redshift $z \sim 1$. We thus expand on \citet[][]{allevato21} by following
their methodology to create realistic mock catalogs of galaxies and active BHs
and (i) we include on top of a regular Gaussian scatter a positive covariant
correlation between the BH and the host galaxy stellar mass at fixed halo
mass; (ii) we focus on redshift $\sim$1.2, i.e.\ close to the peak of the AGN
activity \citep[e.g.][]{shankar09a,madau_dickinson14}; and (iii) we allow the
satellite AGN fraction to vary.  Recently \citet{farahi19} applied a similar
covariant method approach in order to reveal an anti-correlation between
hot and cold baryonic mass in galaxy clusters.  In the context of BH-galaxy
co-evolution, it would expected that a degree of correlation exists between
the properties of the central BH and the galaxy, especially in any scenario
involving an AGN feedback.

The AGN bias as a function of host galaxy stellar mass is thus a crucial
diagnostic of the black hole-galaxy connection which, together with the
bias as a function of black hole mass \citep{shankar20}, can provide
tight constraints on the co-evolution of black holes and galaxies. It
is the ideal time to set out the methodology behind the modelling
of AGN bias as a function of stellar mass based on the large amount
of current and future data. Recent multi-wavelength surveys such as
\textit{XMM}-COSMOS, \textit{Chandra} COSMOS legacy survey, AEGIS have in
fact made it possible to study the AGN environment in terms of host galaxy
properties. In particular, several AGN clustering studies have specifically
considered AGN clustering as a function of the host galaxy stellar mass
\citep[e.g.][]{georgakakis14,mountrichas19,viitanen19,allevato19}. Moreover,
in the next years the synergy among J-PAS, eROSITA, 4MOST, LSST, Euclid, and
JWST will allow us to measure the AGN clustering with high statistics, and
to derive host galaxy stellar mass estimates of moderate-to-high luminosity
AGN up to $z \sim 2$ to measure the large-scale bias as a function of host
galaxy properties of millions of objects at different $z$.

The paper is organized as follows. In Section~\ref{sec:methodology}
we will detail the methodology for estimating the covariant scatter
and the semi-empirical model to build our AGN mock catalogues. In
Section~\ref{sec:results} we outline the results, while in
sections~\ref{sec:discussion} and~\ref{sec:conclusions} we present the
discussion and conclusions of this work.

\section{Methodology}%
\label{sec:methodology}

\begin{figure*}%
  \includegraphics[width=\figurewidthtwo]{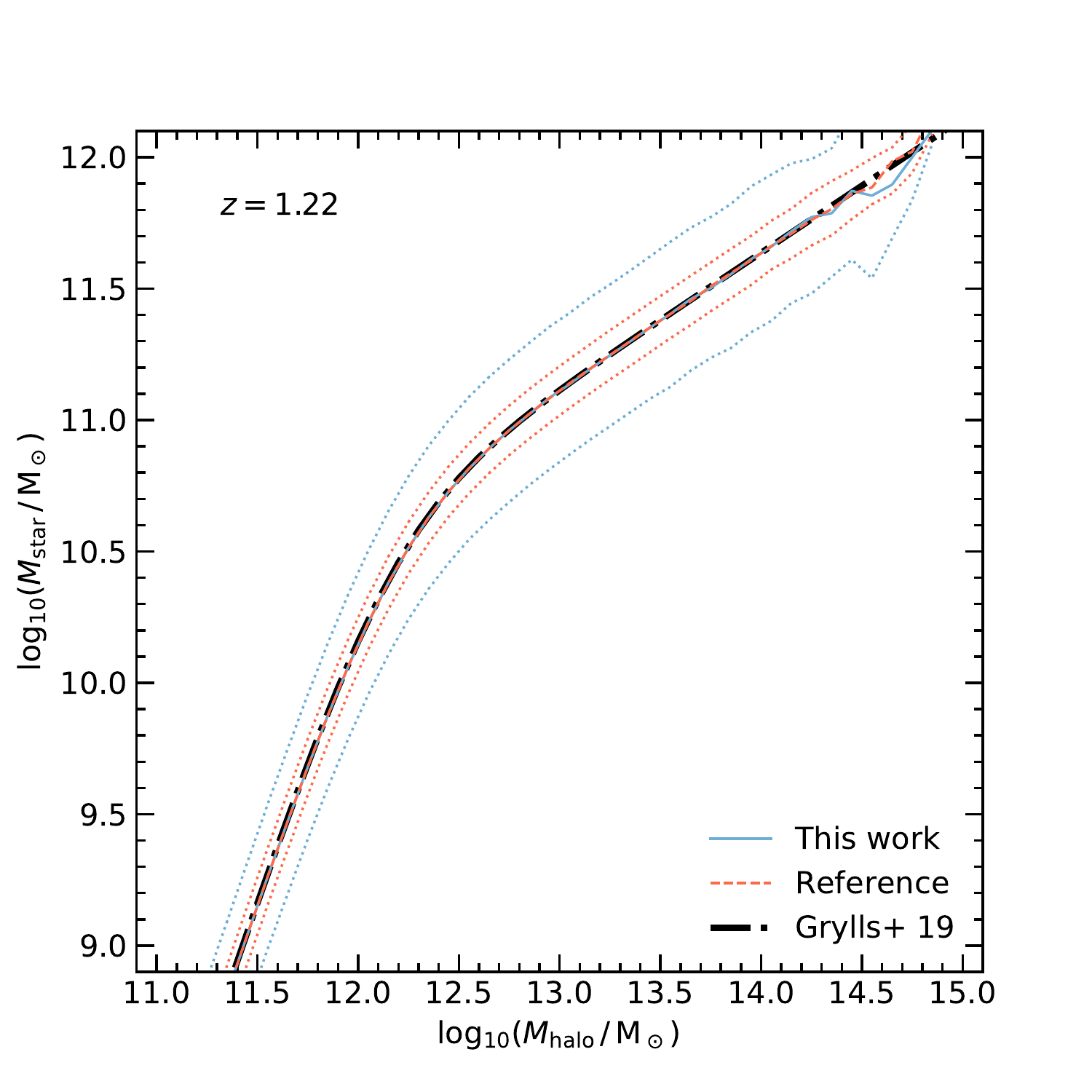}%
  \includegraphics[width=\figurewidthtwo]{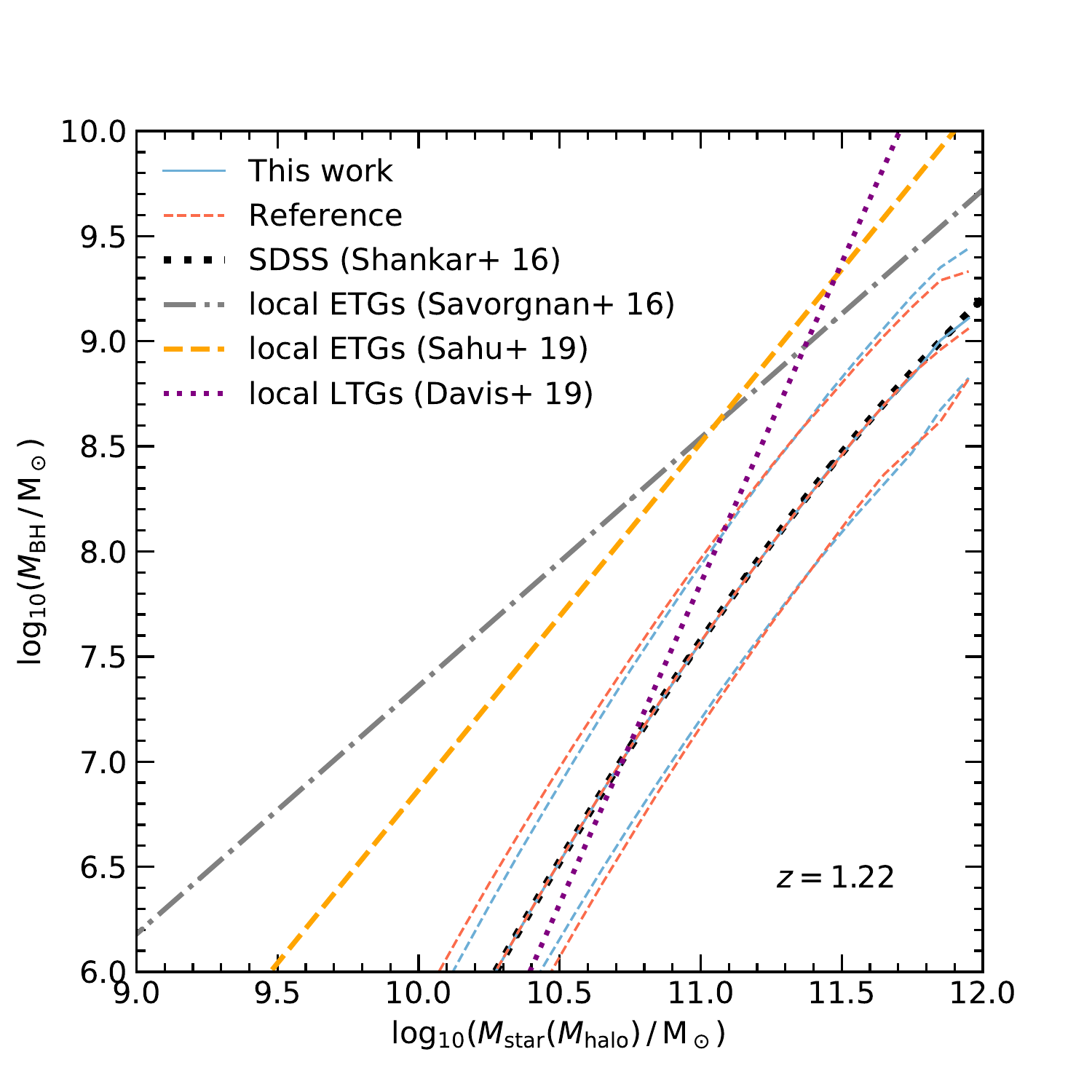}
  \caption{%
    Input scaling relation used to populate DM halos with (active) BHs by
    assigning host galaxy stellar masses $M_\mathrm{star}$ based on given
    DM halo mass $M_\mathrm{halo}$ (left), and BH masses $M_\mathrm{BH}$ to
    given $M_\mathrm{star}$ (right).  The mean scaling relations are shown
    as black dash-dotted lines, while two different methods for the scatter
    in the relations are shown as blue solid (this work, covariant scatter)
    and red dashed (reference) lines. The width of the distributions ($20\%$
    and $80\%$ quantiles) are shown above and below the relations using the
    same colors and dotted lines.
  }%
  \label{fig:mhalomstarmbh}
\end{figure*}

In this work we will investigate the large-scale clustering properties of
X-ray selected AGN, and moreover the imprint of covariant scatter between
the central black hole mass ${M_\mathrm{BH}}$ and the host galaxy stellar
mass ${M_\mathrm{star}}$. For this purpose, we create mock catalogs of
active and non-active galaxies by using semi-empirical models (SEMs) based
on large N-body simulations, and measure the large-scale bias as a function
of stellar mass $b({M_\mathrm{star}})$ as well as the projected two-point
correlation function $w_\mathrm{p}(r_\mathrm{p})$ of mock AGNs.  The full
description of numerical routines to create mock catalogs of galaxies and
their BHs using SEMs is given in \citet{allevato21} at $z \sim 0.1$.

We extend this work to higher redshift $z = 1.22$, and investigate the role of
covariant scatter in assigning ${M_\mathrm{star}}$ and ${M_\mathrm{BH}}$ to
mock AGNs. We only describe the important steps in the generation of the AGN
mock catalogs and we refer the reader to \citet{allevato21} for the details.

We note that our results are specific to the redshift of interest, starting
from the input DM halo catalog and input scaling relations which evolve with
redshift \citep[e.g.][]{grylls19}, and upon assigning large-scale bias to
DM halos $b = b(M_\mathrm{halo}, z)$, with bias increasing with $z$ at a
given $M_\mathrm{halo}$ \citep{sheth01,vandenbosch02}.

\subsection{Input Scaling Relations}%
\label{sub:agn_mock_catalogs}

\begin{figure*}%
  \includegraphics[width=\figurewidthtwo]{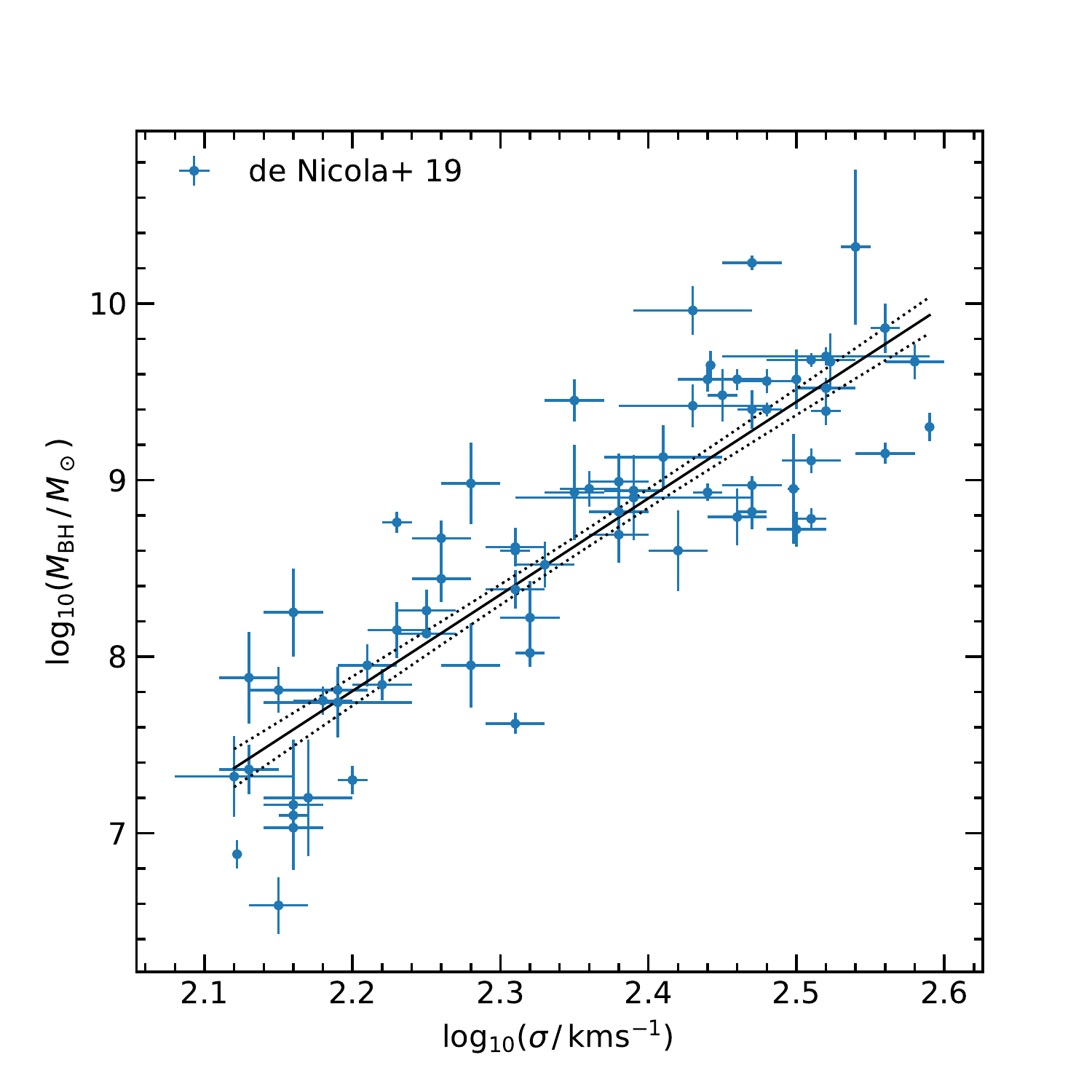}%
  \includegraphics[width=\figurewidthtwo]{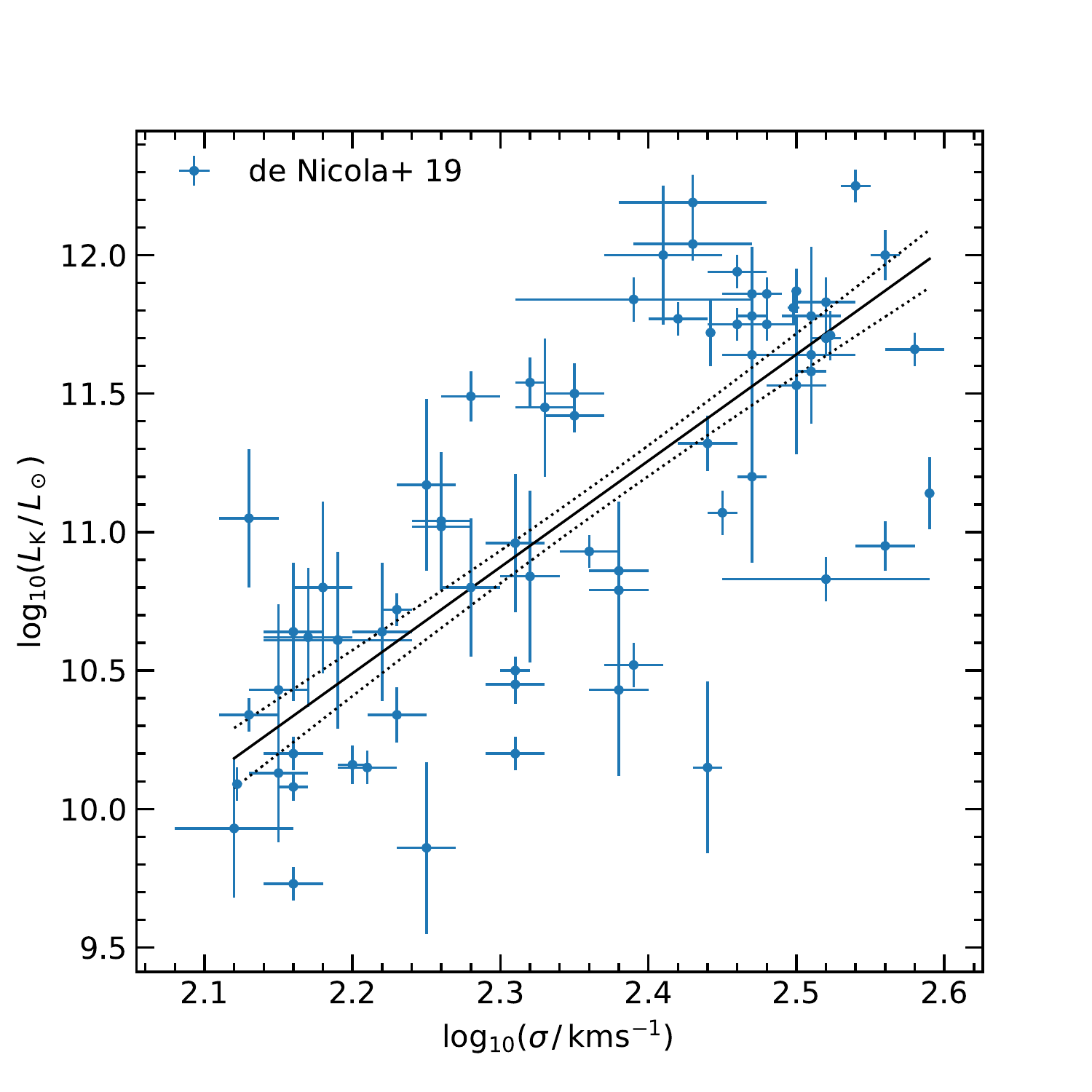}
  \caption{%
    The local galaxy data and the best-fit scaling relations used to derive
    the covariance matrix of the scatter for `This work'. Left and right
    panels show the velocity dispersion against black hole mass, and K-band
    luminosity, respectively. The dashed black lines correspond to $16\%$ and
    $84\%$ quantiles of the best-fit relations based on $20\,000$ Monte Carlo
    samples.%
  }%
  \label{fig:scaling_sigma}
\end{figure*}

For building mock catalogs of AGNs, we start from the dark matter (DM)
halo catalog extracted from the MultiDark Planck 2 (MDPL2) simulation
\citep{riebe13,klypin16} at redshift $z=1.22$, close to the redshift of recent
X-ray AGN clustering measurements \citep[e.g.][]{viitanen19,allevato19}.
The MDPL2 simulation is a cosmological DM only N-body simulation, with a box
size of 1 $\,\mathrm{Gpc}/h$, $3\,840^3$ DM particles, and a mass resolution
of $1.51 \times 10^9 \,\mathrm{M}_\odot/h$.  The cosmology used in the
simulation is flat $\Lambda$CDM with $h = 0.6777$, $\sigma_8 = 0.8228$, and
$\Omega_m = 0.307115$. We use the ROCKSTAR \citep[][]{behroozi13a} DM halo
catalog.The catalogs contain both central/parent halos and satellite halos with
unstripped mass at infall, and the 3-dimensional positions for estimating the
projected two-point correlation function $w_\mathrm{p}$.

In the semi-empirical approach, each halo is assigned a galaxy and central BH
(active or not) following the most up-to-date empirical relations. This
approach has the benefit of avoiding the need to model physical AGN processes,
while reproducing the galaxy/BH statistical properties, such as stellar mass
function and luminosity function. First, halos and subhalos are populated with
stellar masses ${M_\mathrm{star}}$ according to a $M_\mathrm{star} =
M_\mathrm{star}(M_\mathrm{halo}, z)$ (we drop the explicit $z$ dependence and
focus on $z = 1.22$ from now on) relation using the \citet[][]{moster10}
formulae, and the recent updated parameters of \citet{grylls19}.

Galaxies are then populated with BHs according to an input
${M_\mathrm{BH}}({M_\mathrm{star}})$ relation as derived in equation 6
of~\cite{shankar16}, inclusive of the scatter. This relation is significantly
lower in normalization and steeper than relations inferred for BHs with
dynamically measured masses \citep[e.g.][]{kormendy_ho13}. In this work we will
also explore how different input scaling relations (including relations derived
for early-type (ET) and late-type (LT) galaxies) affect AGN clustering
measurements as a function of the host galaxy ${M_\mathrm{star}}$. It is worth
noticing that we assume that all BHs share the same probability of residing in
ET and LT galaxies, as well as star-forming and quenched galaxies. In
Sec.~\ref{sec:results} we will show the effect of relaxing this assumption.

As we investigate in this work, the scatter in the input
${M_\mathrm{star}}({M_\mathrm{halo}})$ and ${M_\mathrm{BH}}({M_\mathrm{star}})$
relations prove to be important parameters in dictating the clustering
properties of mock AGN\@. For this purpose, we will use two different
prescriptions for the scatter, which we label \emph{This work}, and
\emph{Reference} hereafter. The two methods differ in the magnitude of the
scatter in the input ${M_\mathrm{star}}({M_\mathrm{halo}})$ and
${M_\mathrm{BH}}({M_\mathrm{star}})$ relations, and in the case of covariant
scatter we also explicitly define the covariance of the scatter between
${M_\mathrm{star}}$ and ${M_\mathrm{BH}}$ for a given DM halo mass
${M_\mathrm{halo}}$, as detailed further in
Section~\ref{sub:local_scaling_relations}.

Figure~\ref{fig:mhalomstarmbh} shows the ${M_\mathrm{star}}({M_\mathrm{halo}})$
and ${M_\mathrm{star}}({M_\mathrm{BH}})$ relations used in the creation of AGN
mock catalogs, when including the covariant scatter (\emph{This work}) and the
\emph{Reference} case. For comparison, in the right panel of
Fig.\ref{fig:mhalomstarmbh} we show the comparable scaling relations of local
galaxy samples with dynamically measured black hole masses of
\citet{savorgnan16} (Early-type galaxies; ETGs),~\cite{davis19} (Late-type
galaxies; LTGs), and \citet{sahu19} (ETGs).

To each BH is then assigned an Eddington ratio (and then an X-ray luminosity)
following a probability distribution described by a Schechter function as
suggested in \citet{bongiorno12,bongiorno16,aird17,georgakakis17}. The
Schechter function is defined by two free parameters, the knee of the function
$\lambda_{\star}$, and the index $\alpha$ which describes the power-law
behaviour at Eddington ratio values below the knee.

Each BH can be active or not according to an observationally deduced duty cycle
\citep[][]{schulze15}, i.e.\ a probability for each BH of being active.
Following \citet[][]{schulze15}, for the high-end of the black hole mass
function (above $10^6$ solar masses), the AGN duty cycle decreases with the BH
mass at small redshifts and becomes almost constant at $z>1$.
Figure~\ref{fig:functions} (left panel) shows the X-ray luminosity functions of
our mock AGNs compared to data as derived in \citet{miyaji15} at $z=1.098$. The
lines mark the contribution of AGN in different bins of host galaxy
${M_\mathrm{star}}$. Moreover, the AGN host galaxy stellar mass function in
X-ray luminosity bins is shown in the right panel.

We find that a Schechter Eddington ratio distribution with input parameters
$\log \lambda_\star = -0.6$ and $\alpha = +1.2$ (both parameters are
dimensionless) we are able to reproduce the AGN X-ray luminosity function from
\citet[][]{miyaji15}. We have verified that our results are robust to small
changes in the input Eddington ratio parameters. Further dependencies on the
input Eddington ratio are discussed in more detail in \citet{allevato21}.%

Finally, we follow \citet{allevato19} and assign an obscuration value
$N_\mathrm{H}$ to each BH based on AGN luminosity following \citet{ueda14}.

\subsection{Covariant Scatter}%
\label{sub:local_scaling_relations}

\begin{figure*}%
  \includegraphics[width=\figurewidthtwo]{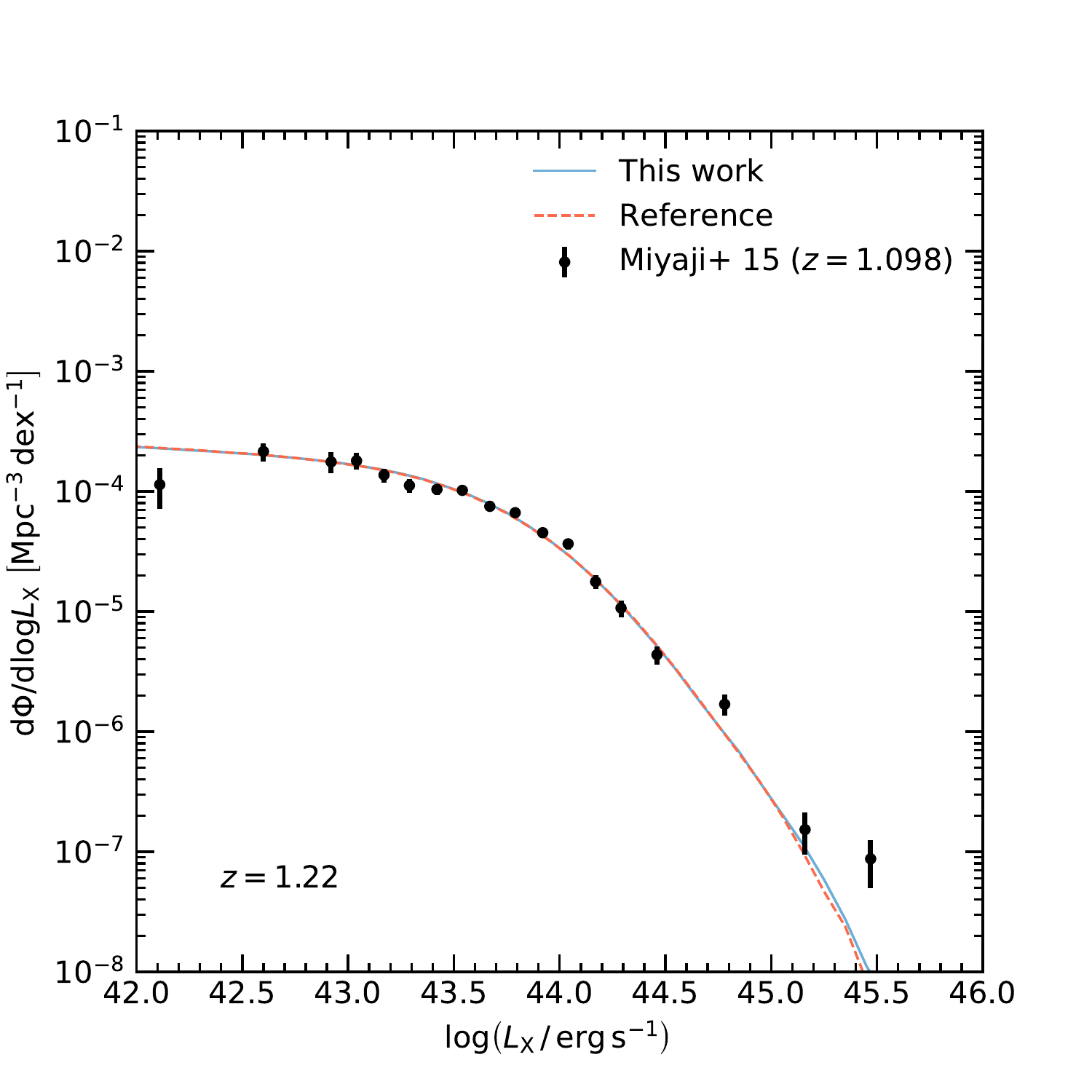}%
  \includegraphics[width=\figurewidthtwo]{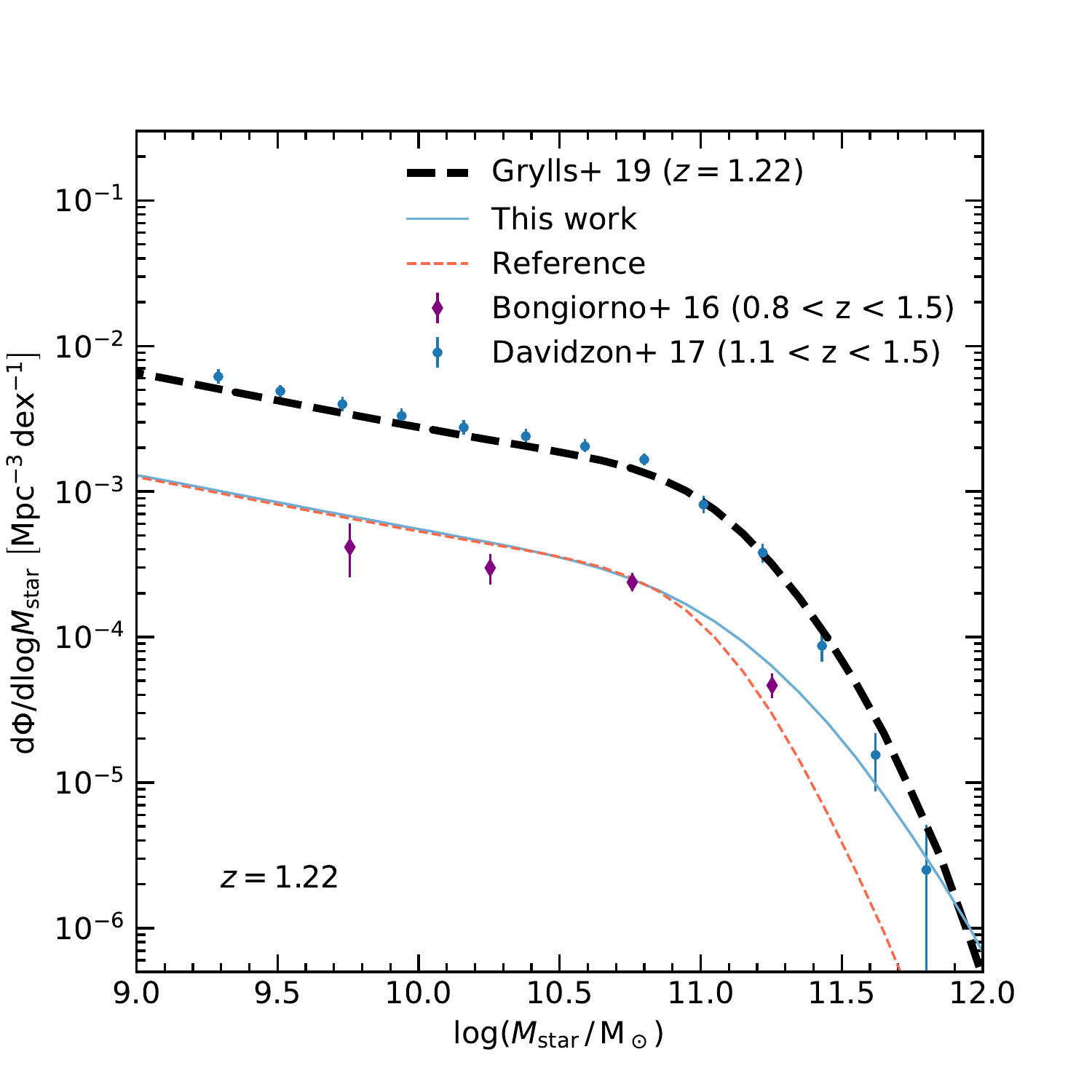}
  \caption{%
    X-ray luminosity (left panel) and host galaxy stellar mass (right panel)
    functions of mock AGNs compared to data. Blue solid lines show the mock AGN
    X-ray (left) stellar mass (right) functions using our covariant scatter
    method i.e.\ \emph{This work}, while red dashed lines show the functions
    using the \emph{reference} method. In the right-hand side panel, we also
    show the mock non-active galaxy stellar mass function as a dashed line and
    for comparison the COSMOS galaxy/AGN host galaxy stellar mass functions of
    \citet[][blue points]{davidzon17} \citet[][purple diamonds]{bongiorno16}.
    The AGN X-ray luminosity function at comparable redshift from
    \citet[][]{miyaji15} (black points), are shown for reference.%
  }%
  \label{fig:functions}
\end{figure*}

In this work we implement two different methods to include the scatter in the
input scaling relations. In the \emph{Reference} case, for a
given halo, we first assign the stellar mass following the ${M_\mathrm{star}} -
{M_\mathrm{halo}}$ inclusive of the scatter. Then, we use the scattered
galaxy stellar mass to assign the black hole mass given
${M_\mathrm{BH}}({M_\mathrm{star}})$
\begin{align}
  \log {M_\mathrm{star}}
  &= \overline{\log M_\mathrm{star}}({\log M_\mathrm{halo}}) + \delta \log {M_\mathrm{star}} \\
  \log {M_\mathrm{BH}} &= \overline{\log M_\mathrm{BH}}({\log M_\mathrm{star}}) + \delta \log {M_\mathrm{BH}},
\end{align}
where $\delta \log {M_\mathrm{star}}$ is a Gaussian random variable with zero
mean and a scatter of $0.11$ \citep[][]{grylls19} and $\delta \log
{M_\mathrm{BH}}$ is a Gaussian random variable with ${M_\mathrm{star}}$
dependent intrinsic scatter of $0.32 - 0.1 \times [\log ({M_\mathrm{star}} /
M_\odot) - 12]$ as suggested by \citet[][]{shankar16}.

However, in the second approach (covariant scatter method, \emph{This Work}),
for a given ${M_\mathrm{halo}}$, we assign the stellar mass ${M_\mathrm{star}}$ and
black hole mass ${M_\mathrm{BH}}$ using the \emph{mean} scaling relation for
both properties. The multivariate normal scatter in ${M_\mathrm{star}}$ and
${M_\mathrm{BH}}$ is then assigned according to a given covariance matrix:
\begin{align}
  \left[
    \begin{matrix}
      \log{M_\mathrm{star}} \\
      \log{M_\mathrm{BH}}
    \end{matrix}
  \right]
  =
  \left[
    \begin{matrix}
      \overline{\log M_\mathrm{star}}({\log M_\mathrm{halo}}) \\
      \overline{\log M_\mathrm{BH}}(\overline{\log M_\mathrm{star}})
    \end{matrix}
  \right] + \mathcal{N} \left( \mathbf{0}, \mathbf{C} \right),
\end{align}
where $\mathcal{N}$ is the 2-dimensional Gaussian distribution with covariance
matrix $\mathbf{C}$ (with covariances $C$)
\begin{equation}
  \mathbf{C} =
  \left[
    \begin{matrix}
      C({\log {M_\mathrm{star}}}, \log {M_\mathrm{star}})
      & C({\log {M_\mathrm{star}}, \log{M_\mathrm{BH}}}) \\
      C({\log {M_\mathrm{star}}, \log{M_\mathrm{BH}}})
      & C({\log {M_\mathrm{BH}}, \log {M_\mathrm{BH}}})
    \end{matrix}
  \right],
\end{equation}
Since data are sparse in terms of the covariance between ${M_\mathrm{star}}$
and ${M_\mathrm{BH}}$ for a given DM halo mass ${M_\mathrm{halo}}$ for AGNs, we
use a local galaxy sample to derive the full covariance matrix by using the
galaxy velocity dispersion and $K$-band luminosities as proxies of
${M_\mathrm{halo}}$ and ${M_\mathrm{star}}$, respectively.

We use the most updated local galaxy sample from~\cite{denicola19} which
includes $84$ galaxies with black hole masses measured either from stellar
dynamics, gas dynamics, or astrophysical masers, velocity dispersion and K-band
luminosities. In this work our interest lies in the massive end of the DM halo
mass function, and given that the AGN duty cycle is not well constrained for
BHs with masses below $10^6 \,\mathrm{M}_\odot$ \citep[][]{schulze15}, we
de-select potential low-mass systems by including only galaxies with velocity
dispersion $\sigma > 10^{2.1}\,\mathrm{km}/\mathrm{s}$, leaving $71$ galaxies
in our final sample, which we show in the left and right panels of
Fig.~\ref{fig:scaling_sigma}.

Our approach is outlined as follows. We start by re-evaluating the scaling
scatter and the full covariance matrix for the scaling relations of the local
galaxy sample of \citet[][]{denicola19}. We use the K-band luminosity
$L_\mathrm{K}$ and velocity dispersion $\sigma$ as proxies of
${M_\mathrm{star}}$ and ${M_\mathrm{halo}}$, in log-linear scale. We then
estimate the ${M_\mathrm{BH}}$--$\sigma$ and ${L_\mathrm{K}}$--$\sigma$ scaling
relations, from which we calculate the scatter and the covariance of
${M_\mathrm{BH}}$ and ${{M_\mathrm{star}}}$ for a given ${M_\mathrm{halo}}$.

For fitting the scaling relations, we use the Bayesian {\textsc{linmix}} routine
\citep{kelly07}, assuming:
\begin{align}
  {l_\mathrm{K}} + \Delta {l_\mathrm{K}} &= \alpha_{L_\mathrm{K}} + \beta_{L_\mathrm{K}} (s + \Delta s) + \epsilon_{L_\mathrm{K}} \\
  {m_\mathrm{BH}} + \Delta {m_\mathrm{BH}} &= \alpha_{M_\mathrm{BH}} + \beta_{M_\mathrm{BH}} (s + \Delta s) + \epsilon_{M_\mathrm{BH}},
\end{align}
where $l_\mathrm{K} = \log {L_\mathrm{K}}$, $m_\mathrm{BH} = \log
{M_\mathrm{BH}}$, and $s = \log \sigma$.

The observational errors, prefixed by $\Delta$ in the equation, are assumed to
be drawn from a multivariate gaussian distribution with a negligible covariance
between the errors \citep{saglia16,vandenbosch16}. The intrinsic scatters
$\epsilon$ are assumed to be drawn from normal distributions with zero means
and $\sigma^2$ variance. We assume uninformative priors and do not mask the
data, and use $20\,000$ Monte Carlo (MC) samples. We then derive the scatter
(denoted by a prefixed $\delta$) from the log-linear relation for each MC
sample $i=1-20\,000$ and for each observation $j=1-71$ from the log-linear
relations:
\begin{align}
  \delta {l_\mathrm{K,i,j}}
  &= {l_\mathrm{K,j}} - (\alpha_{L_\mathrm{K,i}} + \beta_{L_\mathrm{K,i}} s_j), \\
  \delta {m_\mathrm{BH,i,j}}
  &= {m_\mathrm{BH,j}} - (\alpha_{M_\mathrm{BH,i}} + \beta_{M_\mathrm{BH,i}} s_j).
\end{align}
Finally, we estimate the covariance from the scatter for each of the $i$ sample
as follows \citep{farahi19}:
\begin{align}
  C_i({\log L_\mathrm{K}},  \log {L_\mathrm{K}})  &= \frac{1}{N - 1} \sum_{j=1}^{N} \delta l_\mathrm{K,i,j}^2, \\
  C_i({\log M_\mathrm{BH}}, \log {M_\mathrm{BH}}) &= \frac{1}{N - 1} \sum_{j=1}^{N} \delta m_\mathrm{BH,i,j}^2, \\
  C_i({\log L_\mathrm{K}},  \log {M_\mathrm{BH}}) &= \frac{1}{N - 1} \sum_{j=1}^{N} \delta l_\mathrm{K,i,j} \delta m_\mathrm{BH,i,j},
\end{align}
where $N$ is the number of observations For our local sample of $71$ galaxies,
we report $16\%$, $50\%$, and $84\%$ percentiles of the $20\,000$ MC samples in
Table~\ref{tab:linmix}, as well as Fig.~\ref{fig:scaling_sigma}. In detail, for
the scaling relations we find $\alpha_{M_\mathrm{BH}} = -4.19$,
$\beta_{M_\mathrm{BH}} = 5.45$, $\alpha_{L_\mathrm{K}} = 2.06$ and
$\beta_{L_\mathrm{K}} = 3.83$. For the covariance matrix, we find the $50\%$
percentiles $C(\log {L_\mathrm{K}, \log {L_\mathrm{K}}}) = 0.21$, $C(\log
{M_\mathrm{BH}}, \log {M_\mathrm{BH}}) = 0.19$ and $C(\log {L_\mathrm{K}}, \log
{M_\mathrm{BH}}) = 0.07$, which as discussed previously, are also the values we
utilize as the baseline for the full covariance between ${M_\mathrm{star}}$ and
${M_\mathrm{BH}}$ in the covariant scatter method. However, the covariant
scatter derived here may be considered as an upper limit of the scatter in the
input scaling relations. In building our AGN mock catalogs, we fix an upper
limit to the scatter in the BH - stellar mass relation by requiring that the
high-end tail ($M_\mathrm{star} \ge 10^{11.5} \,M_\odot$) of the galaxy stellar
mass function at $z = 1.22$ is not overproduced. This corresponds to a
reduction of $\sigma_{\log M_\mathrm{star}}^2$ by $50\%$ (see
Fig.~\ref{fig:functions}).

\begin{table}
  \begin{center}
\begin{tabular}{lrrr}
  \toprule
  quantity                   & $16\%$ & $50\%$ & $84\%$ \\
  \midrule
  $\alpha_{L_\mathrm{K}}$   & 1.13 & 2.06 & 3.00 \\
  $\beta_{L_\mathrm{K}}$    & 3.44 & 3.83 & 4.22 \\
  $\sigma^2_{L_\mathrm{K}}$ & 0.14 & 0.17 & 0.20 \\
  \midrule
  $\alpha_{M_\mathrm{BH}}$   & -5.11 & -4.19 & -3.26 \\
  $\beta_{M_\mathrm{BH}}$    &  5.06 &  5.45 &  5.84 \\
  $\sigma^2_{M_\mathrm{BH}}$ &  0.13 &  0.16 &  0.20 \\
  \midrule
  $C\left({\log {L_\mathrm{K}},  \log {L_\mathrm{K}}}\right)$   & 0.20 & 0.21 & 0.21 \\
  $C\left({\log {M_\mathrm{BH}}, \log {M_\mathrm{BH}}}\right)$ & 0.18 & 0.19 & 0.19 \\
  $C\left({\log {L_\mathrm{K}},  \log {M_\mathrm{BH}}}\right)$  & 0.07 & 0.07 & 0.08 \\
  \bottomrule
\end{tabular}
\end{center}
 %chktex 27
  \caption{%
    Results from the \textsc{linmix} routine for the quoted quantities derived
    from $20\,000$ MC samples.%
  }%
\label{tab:linmix}
\end{table}

\subsection{Parameter Q}

We also investigate the effect of increasing the relative probability
of satellite BHs of being active. Formally, we define $Q =
U_\mathrm{sat} / U_\mathrm{cen}$, where $U$ is the duty cycle and
the subscripts `sat' and `cen' refer to satellite and central BHs,
respectively.  We can relate the fraction of central and satellite
active BHs to the total fraction of active BHs $U$ via the relation
\citep{shankar20} $U_\mathrm{cen}({M_\mathrm{BH}}) = U({M_\mathrm{BH}})
N({M_\mathrm{BH}}) / \left[ N_\mathrm{cen}({M_\mathrm{BH}}) + Q
N_\mathrm{sat}({M_\mathrm{BH}}) \right ]$ and $U_\mathrm{sat}({M_\mathrm{BH}})
= Q U_\mathrm{cen}({M_\mathrm{BH}})$, respectively. Here $N$ refers to
the numbers of total/central/satellite BHs. In principle, the parameter
$Q$ may not strictly be a constant but a function of stellar mass and/or
environment. However, in the spirit of keeping a flexible and transparent
semi-empirical approach, we will for simplicity keep $Q$ a constant
\citep{shankar20,allevato21}.

It is important to stress that varying the value of $Q$ only modifies the
relative contributions of central and satellite BHs, but it does not alter the
total duty cycle $U$ and thus it does not spoil the match to the AGN luminosity
function. The parameter $Q$ is related to the fraction of AGN in satellite
halos
$f_\mathrm{sat}^\mathrm{AGN}$ by the relation \citep{shankar20}
\begin{equation}
  Q = \frac{f_\mathrm{sat}^\mathrm{AGN}}{1-f_\mathrm{sat}^\mathrm{AGN}}
  \times
  \frac{1 - f_\mathrm{sat}^\mathrm{BH}}{f_\mathrm{sat}^\mathrm{BH}}
\end{equation}
where $f_\mathrm{sat}^\mathrm{AGN}$ is defined via summation over all DM halos
$i$:
\begin{equation}
  f_\mathrm{sat}^\mathrm{AGN}
  = \frac{\sum_i U_{\mathrm{sat},i}}%
         {\sum_i \left(U_{\mathrm{sat},i} + U_{\mathrm{cen},i}\right)},
\end{equation}
and
\begin{equation}
  f_\mathrm{sat}^\mathrm{BH}
  = \frac{N_\mathrm{sat}}{N_\mathrm{sat} + N_\mathrm{cen}}
\end{equation}
is the total fraction of (active and non active) BHs in satellites. In this
work we will study how the $Q$ parameter affects AGN clustering estimates,
and in particular the AGN $b({M_\mathrm{star}})$.

\subsection{Mock AGN samples}
\label{sub:mock_agn_samples}

For a closer comparison between our semi-empirical models and observations, we
match our AGN simulated mocks to the observed samples in terms of
$L_\mathrm{X}$ and stellar mass distributions. We also add to the intrinsic
scatter in the stellar mass -- halo mass relation, a measurement error of $0.20
\,\mathrm{dex}$ to better reproduce the observed scatter in the stellar mass
estimates in the COSMOS AGN samples \citep[][]{allevato19,viitanen19}. Our
reference observed AGN samples are from XMM/Chandra-COSMOS
\citep[][]{allevato19, viitanen19} and XMM-XXL \citep[][]{mountrichas19}.
\citet[][]{mountrichas19} measured the large-scale bias as a function of host
galaxy stellar mass of $407$ moderate luminosity X-ray selected AGN in the
\textit{XMM}-XXL survey at redshift $0.5 < z < 1.2$ (mean $L_\mathrm{X} =
10^{43.7}\,\mathrm{erg}/\mathrm{s}$). \citet[][]{viitanen19} measured
large-scale bias of $632$ \textit{XMM}-COSMOS moderate luminosity X-ray type 1
\& 2 AGN at $0.1 < z < 2.5$ (mean $z \sim 1.2$) and mean $L_\mathrm{X} =
10^{43.7} \,\mathrm{erg}/\mathrm{s}$. Finally, at similar luminosities,
\citet[][]{allevato19} measured with high accuracy the large-scale bias of
$800$ X-ray AGN in the Chandra-COSMOS Legacy (CCL) Survey at $z \sim 1$.

In the following, we also compare with the AGN semi-empirical model of
\citet[][]{aird21}, for which we apply an X-ray luminosity cut at
${L_\mathrm{X}} > 10^{42}$. Lastly, we compare our predictions with the
clustering estimates of SDSS quasars at $z \sim 1.4$ measured by
\citet[][]{richardson12}, and for this test we impose a limit on luminosity of
${L_\mathrm{X}} > 10^{44} \,\mathrm{erg}\,\mathrm{s}^{-1}$ and on Hydrogen
column density of ${N_\mathrm{H}} < 10^{22} \,\mathrm{cm}^{-2}$ to select only
optical, nominally Type I AGN \citep[e.g.,][]{ricci17}. We list some of the
main features of each of our reference AGN mock subsample in
Table~\ref{tab:mock}.

\begin{table}
  \begin{center}
\begin{tabular}{lrrrr}
\toprule
                sample &    $N_\mathrm{halo}$ &     $N_\mathrm{AGN}$ &   $Q$ & $f_\mathrm{sat}^\mathrm{AGN}$ \\
\midrule
        Aird+ 21 -like &        $5\,746\,229$ &        $1\,022\,468$ &  1.00 &                 0.12 \\
        Aird+ 21 -like &        $5\,768\,258$ &        $1\,026\,768$ &  2.00 &                 0.22 \\
    Allevato+ 19 -like &        $1\,459\,756$ &           $217\,799$ &  1.00 &                 0.10 \\
    Allevato+ 19 -like &        $1\,468\,620$ &           $219\,237$ &  2.00 &                 0.18 \\
    Allevato+ 19 -like &        $1\,454\,717$ &           $216\,930$ &  4.00 &                 0.31 \\
    Allevato+ 19 -like &        $1\,464\,559$ &           $218\,672$ &  6.00 &                 0.40 \\
  Richardson+ 12 -like &           $324\,580$ &            $27\,078$ &  1.00 &                 0.07 \\
  Richardson+ 12 -like &           $321\,187$ &            $26\,654$ &  4.00 &                 0.23 \\
  Richardson+ 12 -like &           $324\,981$ &            $26\,974$ &  6.00 &                 0.31 \\
\bottomrule
\end{tabular}
\end{center}

  \caption{%
    Properties of the AGN sub-samples used in this work. The columns correspond
    to the name of the sample, number of DM halos, number of AGN defined as the
    sum of the duty cycles, $Q$ parameter, and corresponding AGN satellite
    fraction. See the text for the details on the sample definitions.
  }
  \label{tab:mock}
\end{table}

\subsection{Two-point correlation function and large-scale bias}%
\label{sub:two_point_correlation_function}

\begin{figure*}%
  \includegraphics[width=\figurewidthtwo]{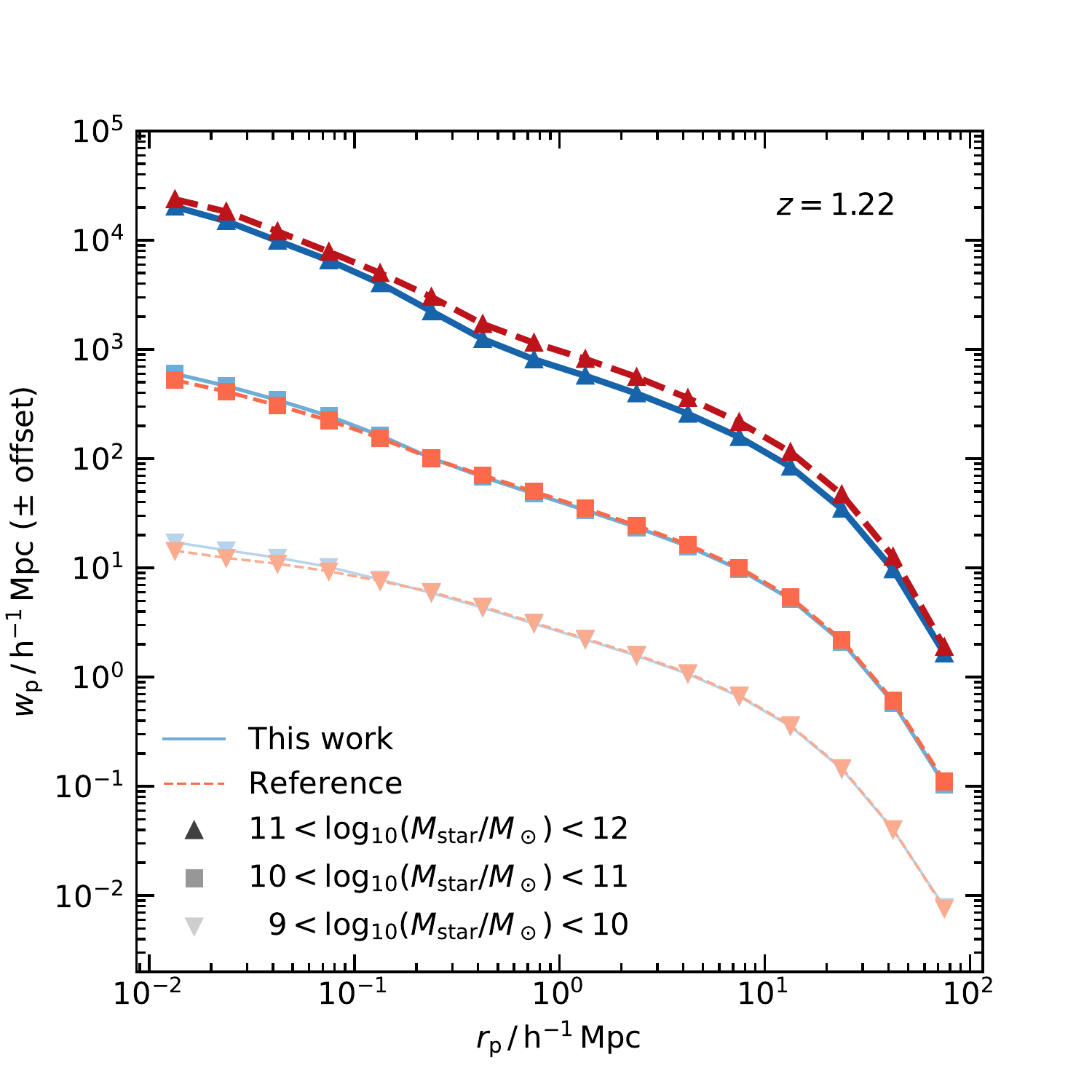}%
  \includegraphics[width=\figurewidthtwo]{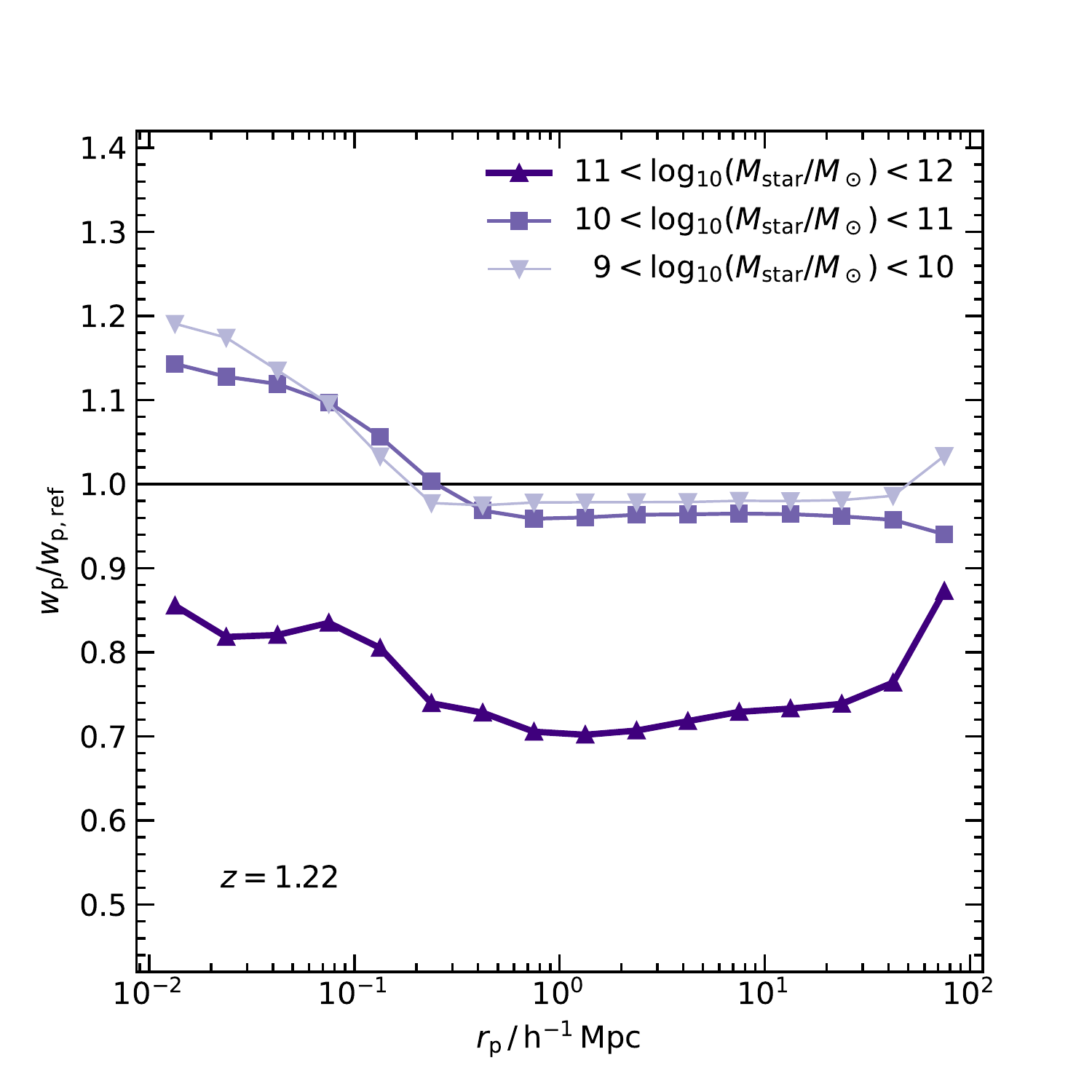}
  \caption{%
    Left: projected correlation functions of mock AGNs
    in bins of host galaxy ${M_\mathrm{star}}$ assuming $Q=1$.
    Blue solid lines show the covariant scatter method, while red dashed lines
    show the reference case. Bins of ${M_\mathrm{star}}$ (in units of
    $\,\mathrm{M}_\odot$) are indicated by the symbols and the shades of color.
    For an easier comparison between the two methods, the lowest and the highest
    ${M_\mathrm{star}}$ bins have been offset in the direction of
    $w_{\mathrm{p}}$ by $-1$ and $+1$ dex, respectively. Right: the relative
    bias in $w_\mathrm{p}(r_\mathrm{p})$ using the two methods for the scatter.
    Different symbols indicate the different stellar mass bins following
    the legend.%
  }%
  \label{fig:wps}
\end{figure*}

For quantifying the clustering of mock AGNs, we use two complementary
approaches. Firstly, we compute the projected two-point correlation function
using the three-dimensional positions of the AGN host DM halos. Secondly,
we calculate the AGN large-scale bias by averaging over the biased parent
DM halo population.

We estimate the projected two-point correlation function of mock AGNs by using
the 3-d positions of the hosting DM halos, following \citet{davis_peebles83}:
\begin{equation}
  w_\mathrm{p}(r_\mathrm{p}) = 2 \times \int_0^{\pi_{\max}} \xi(r_\mathrm{p}, \pi)
  \mathrm{d}\pi
\end{equation}
where $\xi(r_\mathrm{p}, \pi)$ is the 2-dimensional Cartesian two-point
correlation function \citep[e.g.][]{peebles80}. Physical distances
$r_{\mathrm{p}}$ and $\pi$ correspond to the perpendicular and parallel
(defined with respect to a far-away observer) separations, defined for each AGN
pair $DD_{i,j}(r_{\mathrm{p}}, \pi)$ separately. Using periodic boundary
conditions in the simulation box with volume $V = 10^9
\,{({h}^{-1}\mathrm{Mpc})}^3$, $\xi(r_\mathrm{p},\pi)$ is estimated using
\begin{align}
  1 + \xi(r_{\mathrm{p}}, \pi)
  &= \frac{DD(r_{\mathrm{p}},\pi)}{RR(r_{\mathrm{p}}, \pi)}, \\
  DD(r_{\mathrm{p}}, \pi) &= \sum_{i,j} U_i \times U_j, \\
  RR(r_{\mathrm{p}}, \pi) &\approx \frac{\Delta V}{V} {\left(\sum_k U_k\right)}^2
\end{align}
where $DD$ is the sum of all unique mock AGN pairs $i,j$ within cylindrical
volume element $\Delta V$ defined as the volume enclosed by $\log
r_{\mathrm{p}} \pm \Delta \log r_{\mathrm{p}}/2$ and $\pi \pm \Delta \pi/2$
weighted by the AGN duty cycle $U$, and $RR$ is the expected number of randomly
distributed mock AGN pairs within the same volume
\citep[e.g.][]{alonso12,sinha20}. For estimating the correlation function, we
use $r_\mathrm{p}=0.1-100\,{h}^{-1}\mathrm{Mpc}$, and $\pi=0-40$ ($\pi_{\max} =
40 \,{h}^{-1}\mathrm{Mpc}$) with bin sizes $\Delta \log
(r_\mathrm{p}/\,{h}^{-1}\mathrm{Mpc}) = 0.25$ and $\Delta \pi = 1
\,{h}^{-1}\mathrm{Mpc}$, respectively. We use the publicly available
\textsc{CorrFunc} code~\citep{sinha20}.

For the large-scale bias $b({M_\mathrm{star}})$ we estimate the bias for each
DM halo separately. Halos are labeled as central or satellite. For each central
halo, we assign a value of the large-scale bias according
to~\cite{vandenbosch02,sheth01}. For each satellite DM halo we assign a
large-scale bias value based on the mass of its parent halo, as each satellite
traces the dense environment of the parent halo%
\footnote{%
  We further verify that the DM halo bias of~\cite{sheth01} is consistent
  (within 10\%) with the large-scale bias estimated through the one-dimensional
  2pcf $b = \sqrt{\xi(r)/\xi_\mathrm{DM}(r)}$, by directly calculating $\xi(r)$
  for MDPL2 halos in several narrow mass bins and estimating $\xi_\mathrm{DM}$
  \citep{eisenstein_hu98}.%
}.
Then, we follow the formalism of \citet[][]{shankar20,allevato21} to derive the
bias of mock AGN and normal galaxies as a function of galaxy stellar mass, by
using the $Q$ parameter. The bias of mock objects with stellar mass in the
range $M_\mathrm{star} \pm \mathrm{d}M_\mathrm{star}/2$ is estimated as a
weighted average:
\begin{equation}
\begin{split}
  b({M_\mathrm{star}})
  &      = \left[ \sum_{i=1}^{N_\mathrm{cen}} U_{\mathrm{cen},i}(M_\mathrm{star}) b_{\mathrm{cen},i}(M_\mathrm{star}) \right. \\
  &      + \left. \sum_{i=1}^{N_\mathrm{sat}} U_{\mathrm{sat},i}(M_\mathrm{star}) b_{\mathrm{sat},i}(M_\mathrm{star}) \right] \\
  & \left/ \left[ \sum_{i=1}^{N_\mathrm{cen}} U_{\mathrm{cen},i}(M_\mathrm{star}) \right. \right.
  + \left. \sum_{i=1}^{N_\mathrm{sat}} U_{\mathrm{sat},i}(M_\mathrm{star}) \right]
\end{split}
\end{equation}
When $Q = 1$ and $U_\mathrm{sat} = U_\mathrm{cen}$, then on average central and
satellite BHs share equal probabilities of being active.

\section{Results}%
\label{sec:results}

In this section we present our main results in terms of projected two-point
correlation function $w_{\mathrm{p}}(r_{\mathrm{p}})$, AGN large-scale bias as
a function of galaxy stellar mass $b({M_\mathrm{star}}$) and mean AGN halo
occupation distribution (HOD) $\left\langle N(M_\mathrm{halo}) \right\rangle$,
i.e.\ the average number of AGN as a function of the halo mass. In what
follows, we label as ``\emph{This work}'' all results based on the covariant
scatter method, while we use the ``\emph{Reference}'' label for all models
characterised by independent Gaussian scatters in all the input scaling
relations.

The left panel of Fig.~\ref{fig:wps} shows the projected correlation function
$w_{\mathrm{p}}(r_{\mathrm{p}})$ of mock AGNs created using the covariant
scatter method (solid lines) and the reference model (dashed red lines), for
different host galaxy stellar mass bins. The right panel of Fig.~\ref{fig:wps}
plots instead the relative bias between the two different approaches. On large
scales, the difference is in the highest stellar mass bin $11 <
\log{(M_\mathrm{star}/M_\odot)} < 12$, where the clustering strength of the
covariant scatter is a factor of $\sim 0.7$ lower compared to the reference
case. At smaller stellar masses, the two cases differ by a more moderate factor
of $\lesssim 0.9$. Meanwhile, at lower scales $r_{\mathrm{p}}$, and for the
lower stellar mass bins, we find an opposite trend, where the clustering
strength of the covariant scatter case is higher by up to a factor of $\sim
1.2$.

\begin{figure}%
  \centering
  \includegraphics[width=\linewidth]{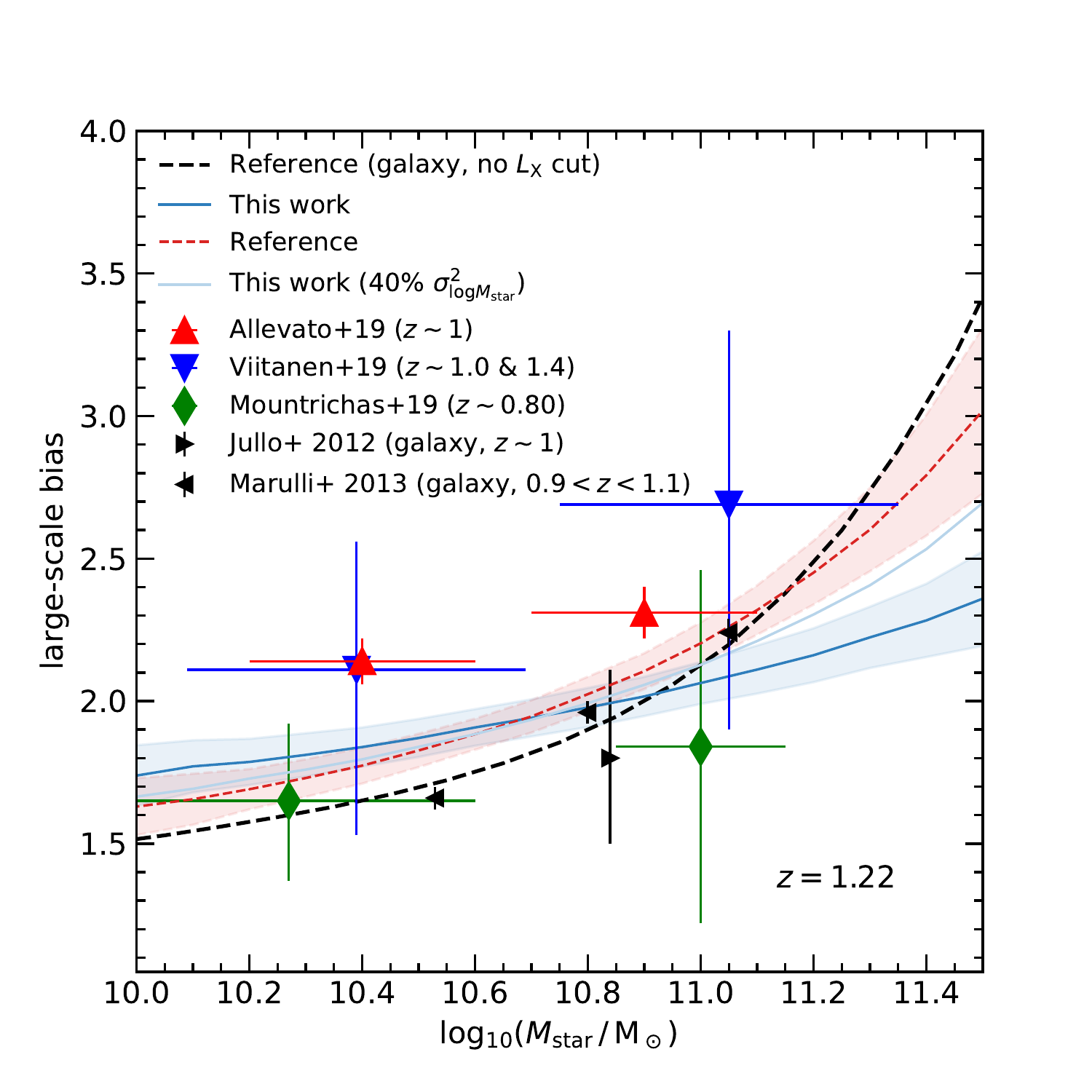}%
  \caption{%
    Large-scale bias as a function of the galaxy stellar mass of normal
    galaxies (black dashed line) compared to measurements of
    \citet[][]{jullo12} and \citet[][]{marulli13} (black left and right
    triangles); and mock AGNs when using the covariant scatter method, as solid
    (`This work') and dashed (`Reference') lines, respectively. The shaded
    region shows the 1$\sigma$ error of the AGN bias using a COSMOS-like volume
    (see the text for the details). The lighter blue line shows the effect of
    reducing the variance of the stellar mass for a given DM halo mass in the
    covariant scatter method to $40\%$ of the covariant scatter value. The
    markers show the X-ray AGN bias measurements from recent X-ray AGN
    clustering studies at similar redshifts
    \citep[][]{allevato19,mountrichas19,viitanen19} in accordance with the
    legend.%
  }%
  \label{fig:bias_error}
\end{figure}

In Fig.~\ref{fig:bias_error} we show the bias in bins of stellar mass of mock
AGNs for the covariant scatter case (solid blue line), and the reference case
(dashed red line), and for X-ray selected AGN at $z \sim 1$
\citep{allevato19,mountrichas19,viitanen19}.  For these comparisons, we match
the mock AGN sample in terms of X-ray luminosity to the Chandra-COSMOS sample
\citep{allevato19}, as explained in Sec.~\ref{sub:mock_agn_samples}. In the
covariant scatter case, the predicted AGN bias is almost constant as a function
of stellar mass, as the covariance by design tends to naturally increase the
scatter in stellar mass at fixed DM halo mass. In fact, further decreasing the
intrinsic scatter in stellar mass to $40\%$ of the value derived in the
covariant scatter method, leads to an AGN large-scale bias that is closer to
the reference case without covariance (dashed red line). Moreover, we note that
the typical $M_\mathrm{star}$ measurement error in COSMOS (see
Sec.~\ref{sec:methodology}) added on top of the intrinsic scatter bridges some
of the gap in the difference of the biases at $M_\mathrm{star} \gtrsim 10^{11}
\, M_\odot$, where the inclusion of this additional scatter lowers the bias at
a given stellar mass principally in the reference case.

For completeness, we also show the large-scale bias as a function of stellar
mass of VIPERS \citep{marulli13} and COSMOS \citep{jullo12} for all galaxies at
a similar redshift, which is well reproduced by our mock galaxies (i.e., bias
not weighted by the duty cycle and without cuts in the X-ray luminosity, black
dashed line in Fig.~\ref{fig:bias_error}). It is worth noticing that, in the
reference case, normal and active galaxies with the same luminosity cut follow
the same bias - stellar mass relation. A larger degree of discrepancy in the
bias in bins of stellar mass between AGN and normal galaxies is instead
predicted when assuming a covariant scatter, especially at
$\log{(M_\mathrm{star} / M_\odot)} > 11$ (see Sec.~\ref{sec:discussion} for
more discussion).

The predicted bias as a function of stellar mass of mock AGNs is consistent
with the bias estimates of X-ray selected AGN characterized by a large
uncertainty, irrespective of the used method for the scatter. However, the bias
of mock AGNs appears systematically below by at least $\sim 20\%$ the
measurements derived in \citet{allevato19} for Chandra-COSMOS AGN\@.

There could be different, concurrent causes that could explain the offset
between COSMOS and the mocks. The COSMOS field is renown to be characterized by
overdensities at $z < 1$ that might increase the AGN large-scale bias by up to
26\% \citep[e.g.][]{gilli09,mendez16}, and suffers from cosmic variance due to
the small volume. In particular, to estimate the effect of such a finite volume
in our AGN mocks, we provide an estimate of the cosmic variance by sampling the
full MDPL2 simulation box with side length of 1 Gpc/h. We then extract a total
of 36 unique sub-volumes with area $27 \times 27$ ${(\mathrm{Mpc}/h)}^2$
approximately $1.4$ deg$^2$ at $z = 1$, comparable to the COSMOS field
\citep[][] {scoville07}, and length 1 Gpc/h, and calculate the large-scale bias
for each of the 36 sub-boxes individually. We then use the standard deviation
of the large-scale bias from the sub-volumes as an estimate of the effect of
the cosmic variance. The filled areas in Fig.~\ref{fig:bias_error} shows the
$1\sigma$ error of the bias, upon selecting a COSMOS-like volume from the full
MDPL2 simulation box, which is of the order of $\sim$0.1 dex at almost all
stellar masses. Our results would thus imply that, by itself, cosmic variance
cannot account for the full offset between our mocks and the \citet{allevato19}
bias measurements. We will discuss additional factors that could contribute to
the discrepancy between models and data in what follows below. We will
specifically investigate the dependence of the AGN large-scale bias (and
related AGN HOD) on other input parameters, namely the BH - galaxy scaling
relation and the AGN satellite fraction (i.e., the number of AGN in massive
galaxy groups/clusters) as parametrized by $Q$.

\subsection{Dependence on Input Scaling Relations}

\begin{figure}%
  \includegraphics[width=\linewidth]{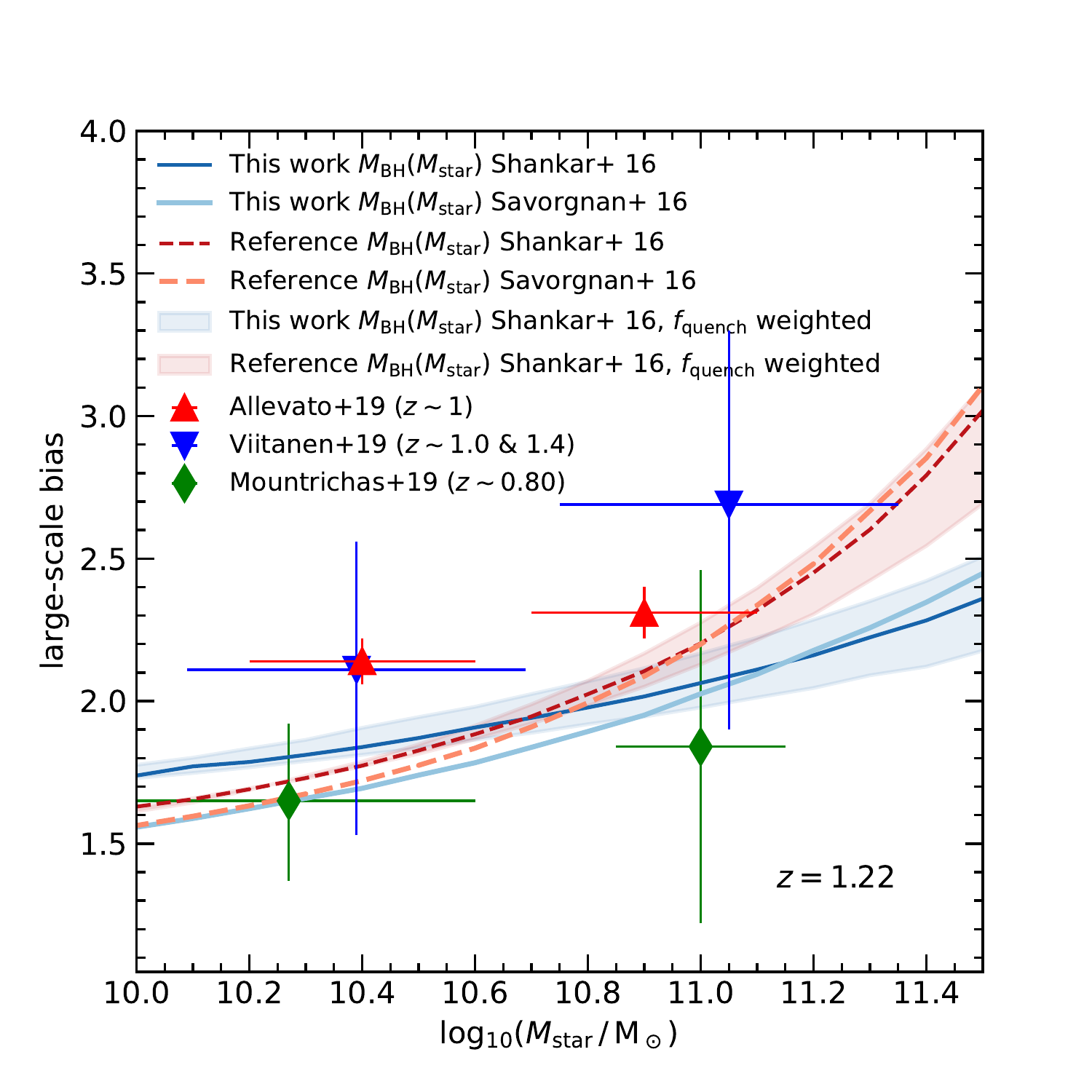}%
  \caption{%
    Large-scale bias as a function of host galaxy stellar mass and its
    dependence on the input ${M_\mathrm{BH}}({M_\mathrm{star}})$ relation. The
    shaded area around \citet{shankar16} relation shows the bias weighted
    additionally by the probability of a galaxy of being quenched as a function
    of the DM halo mass and redshift (assuming $\mu = 2.5$) as given by
    \citet{zanisi21}. The limits correspond to AGN host galaxies consisting
    solely of quenched (upper limit), and non-quenched (lower limit) galaxies.
    Otherwise the symbols and linestyles are the same as in
    Fig.~\ref{fig:bias_error}.
  }%
  \label{fig:bias_mstar_mbh}
\end{figure}

In the creation of AGN mock catalogs we have assigned black hole masses to
galaxies according to their host galaxy stellar mass
following~\cite{shankar16}. Here we explore the impact of varying this
input relation. We adopt the relation derived by \citet[][]{savorgnan16} for a
sample of galaxies with dynamically measured BH masses, as presented in Eq.~3
of \citet[][]{shankar19}, which predicts higher ${M_\mathrm{BH}}$ for a given
${M_\mathrm{star}}$. These two chosen relations broadly bracket the existing
systematic uncertainties in the black hole mass--stellar mass relation of
dynamically measured BHs in the local Universe, with other scaling relations
broadly falling in between them \citep[e.g.,][]{terrazas16,davis19,sahu19}.%

We find that, in the reference case, the AGN large-scale bias as a function of
stellar mass is independent of the particular input BH mass -- stellar mass
relation. When assuming the covariant scatter method instead, the AGN bias
predicted by the \citet{savorgnan16} relation is slightly lower (by $\sim 0.1$
dex) at lower stellar masses as compared to assuming the relation proposed by
\citet[][]{shankar16}.

In our model we also assume that AGNs reside in a mixed population of host
galaxies, i.e.\ active BHs share the same probability of being active in star
forming and quenched galaxies. We can relax this assumption and explore the
effect on the AGN bias as a function of stellar mass of having all AGNs in
quenched galaxies. To this purpose, we assign to each AGN a probability of
being in a quenched host galaxy $f_{quench}$ as a function of halo mass and
redshift following \citet{rodriguez-puebla15} and \citet{zanisi21}. In detail,
we use $f_\mathrm{quench} = 1 / [b_0 + (M(z, \mu) \times 10^{12} /
M_\mathrm{halo})]$, where $M(z, \mu) = M_0 + {(1 + z)}^\mu$, with
$M_\mathrm{halo}$ given in units of solar masses, $M_0=0.68$, $b \sim 1$, and
$\mu \sim 2.5$ (see \citealt{zanisi21} for the details). We note that
variations in $\mu=1-4$ affect the large-scale bias at most at the $\sim 5\%$
level as compared to $\mu=2.5$. We show the result in
Fig.~\ref{fig:bias_mstar_mbh} as a shaded area, where the upper (lower) limit
corresponds to the AGN bias in bins of stellar mass when assuming all AGNs in
quenched (non-quenched) galaxies. In particular, assuming that all mock AGNs
reside in quenched host galaxies, which live preferentially in more massive and
biased halos, produces a large-scale bias as a function of stellar mass
slightly higher ($1.05$ times) than when considering AGN in a mixed
population.

\subsection{Dependence on Q}

\begin{figure*}%
  \centering
  \includegraphics[width=\figurewidthtwo]{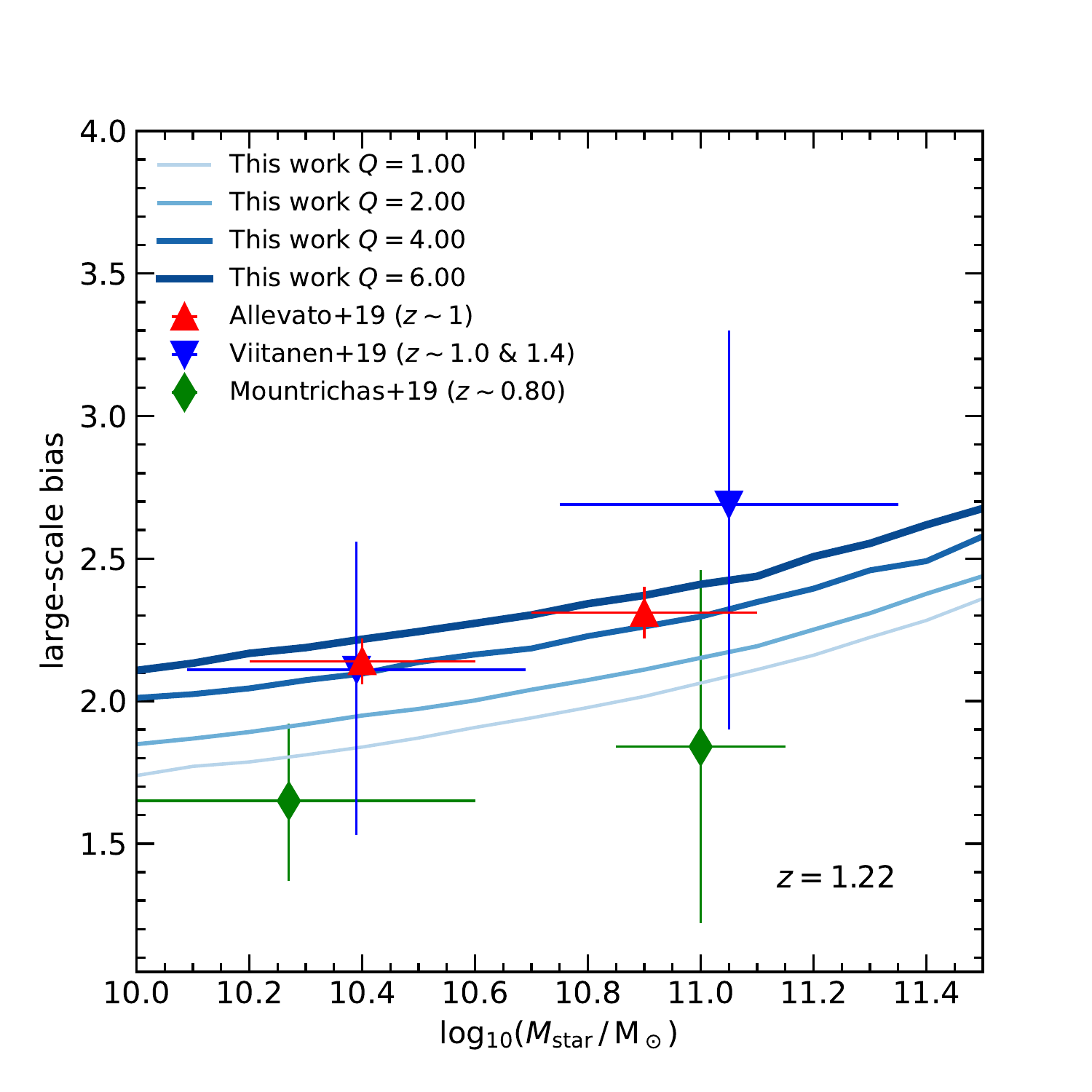}%
  \includegraphics[width=\figurewidthtwo]{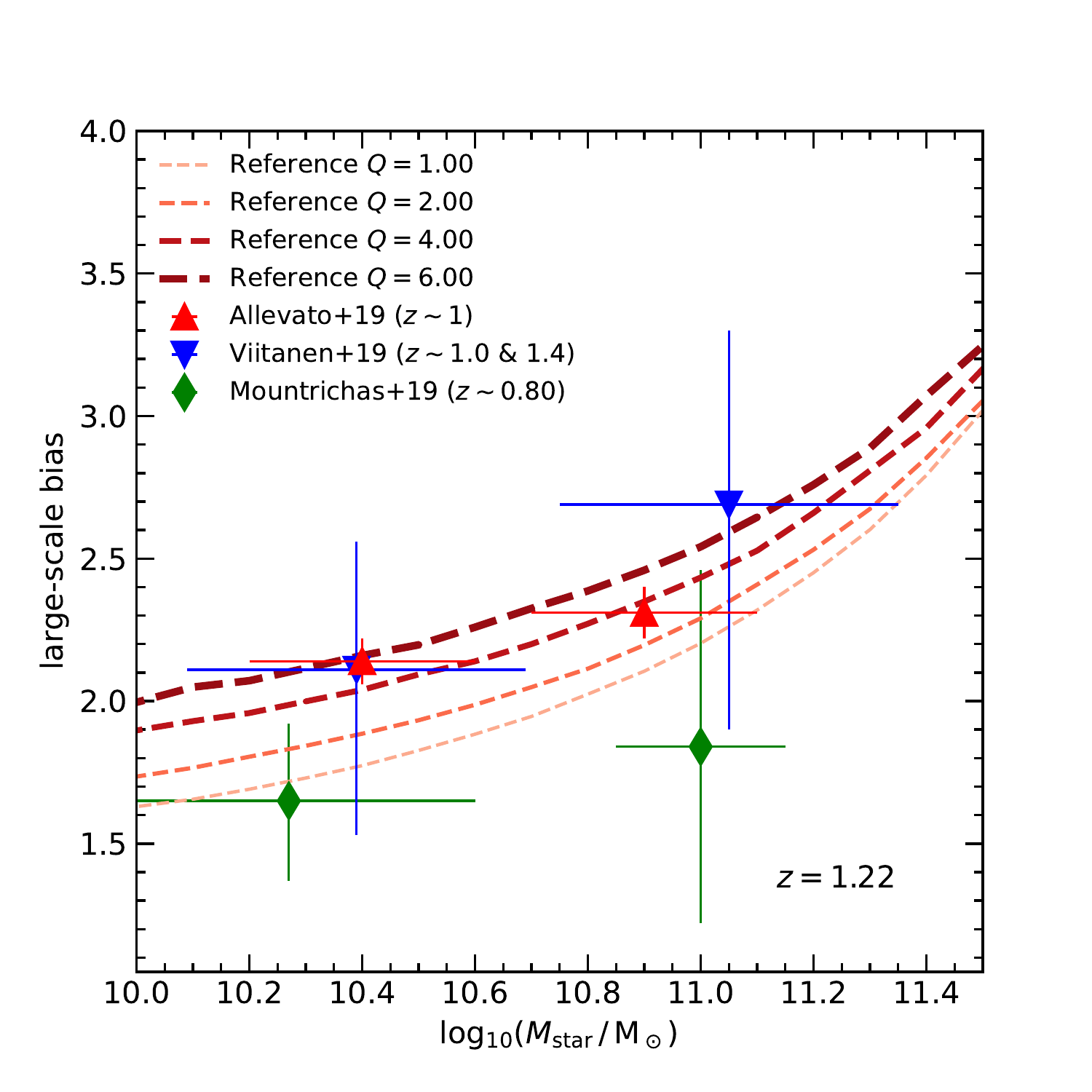}
  \caption{%
    Large-scale bias as a function of stellar mass for different AGN satellite
    fractions parametrized by the $Q$ parameter. In the left (right) panel,
    different lines show the bias using the covariant scatter (reference)
    method with $Q=1-6$ as indicated by different line widths and colors.
    }
  \label{fig:bias_q}
\end{figure*}

In this section we study the effect of varying the relative number of satellite
AGNs parameterized by $Q$. Fig.~\ref{fig:bias_q} shows that higher values of
$Q$ tend to increase the normalization of the bias at all considered stellar
masses, both in \textit{This Work} and in the \textit{Reference} case. This can
be understood as an increase in the fraction of AGNs in highly biased massive
systems, which are likely to host satellite AGNs in the first place. Increasing
the probability of a satellite BHs being active thus increases the bias. We
show this effect for moderate values of $Q=1,2$ (i.e.\ satellite BHs are $1-2$
times more likely to be active than centrals), but also more extreme values of
$Q=4,6$. Bias estimates for X-ray selected COSMOS AGN seem to favour values of
$Q\sim4$, while \emph{XMM}-XXL AGN are more in agreement with models with $Q =
1$.

\begin{figure}%
  \centering
  \includegraphics[width=\linewidth]{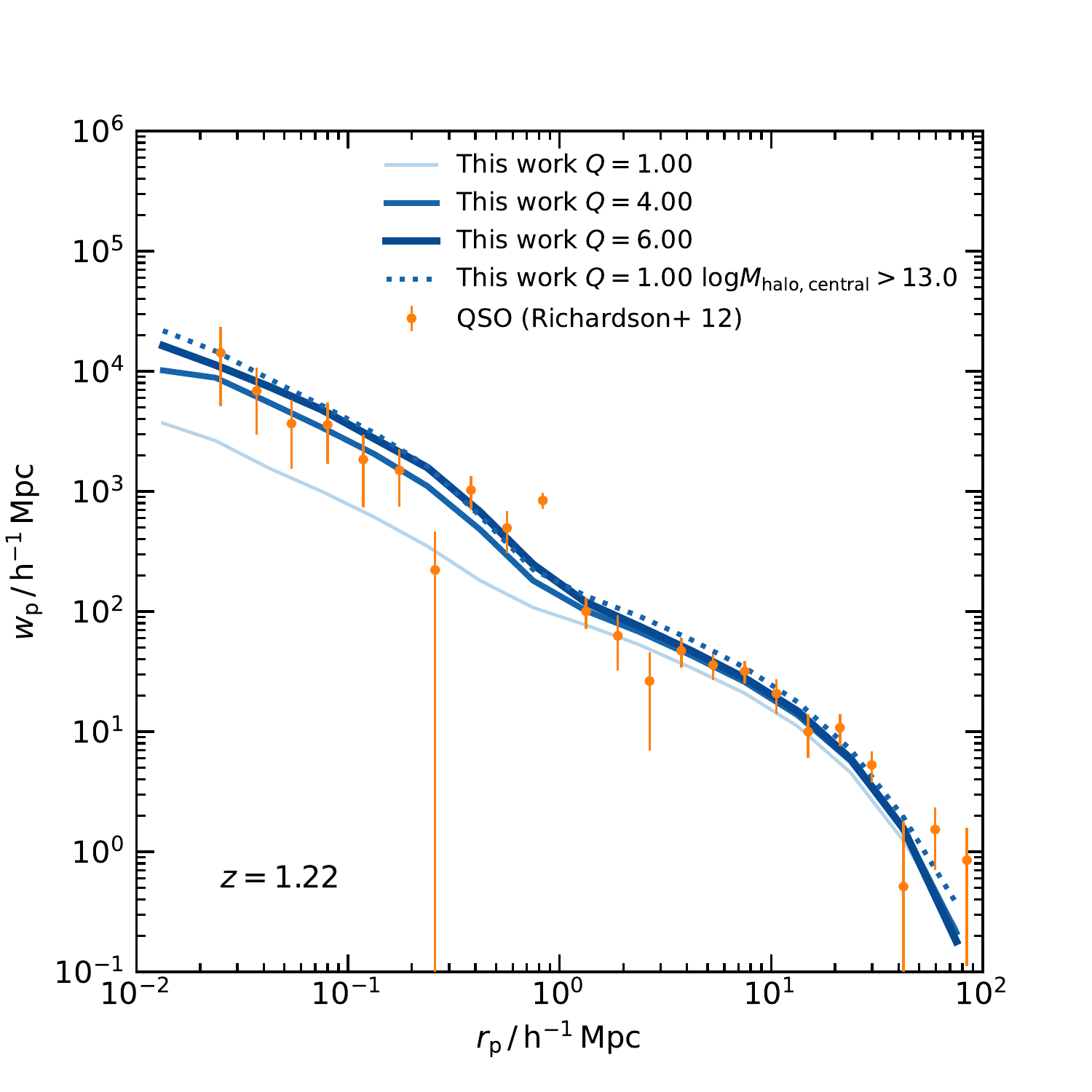}
  \caption{%
    Projected correlation function including the 1-halo term ($r_{\mathrm{p}}
    \le 1-2\,{h}^{-1}\mathrm{Mpc}$) of the QSO-like mock AGN sample
    (${L_\mathrm{X}} > 10^{44} \,\mathrm{erg}\,\mathrm{s}^{-1}$,
    ${N_\mathrm{H}} < 10^{22} \,\mathrm{cm}^{-2}$) compared to
    \citet[][]{richardson12} SDSS QSOs at a similar redshift. The results are
    shown for the covariant scatter method, although there is no significant
    dependence on the method used.
  }%
  \label{fig:gatti_wp}
\end{figure}

\begin{figure}%
  \centering
  \includegraphics[width=\linewidth]{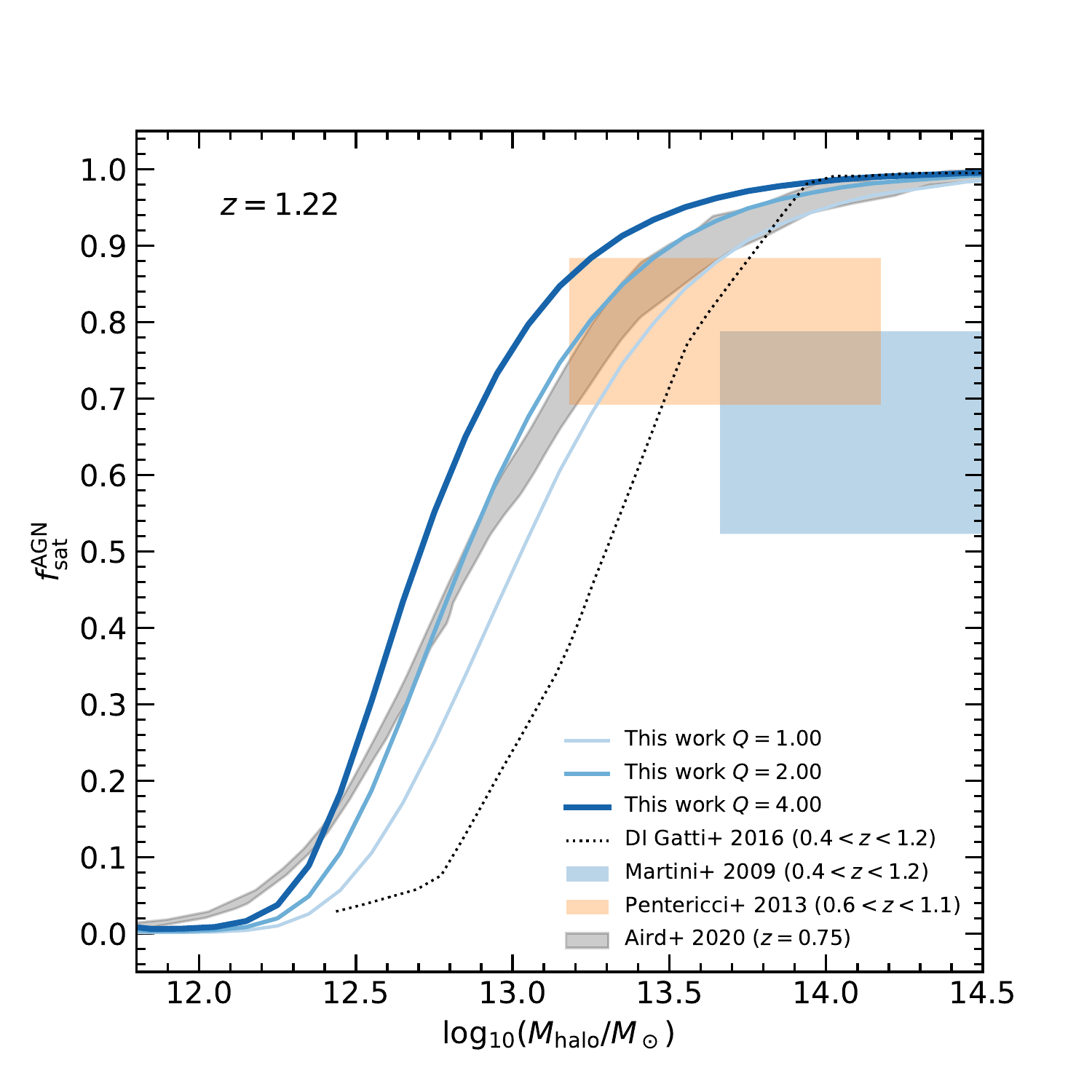}
  \caption{%
      Fraction of satellite AGN as a function of parent host DM halo mass for
      different values of $Q$ parameter. Colored lines with increasing widths
      correspond to ascending values of $Q$ for our mock AGNs, while the shaded
      region shows the limiting values of \citet[][]{aird21} model. The
      semi-analytic model prediction of \citet[][]{gatti16} based AGN
      triggering through disc instabilities (DI) is shown as a dark dotted
      line. \citet[][]{martini09,pentericci13} show measurements of X-ray AGNs
      in galaxy groups/clusters.
    }%
  \label{fig:mhalo_fsat}
\end{figure}

We can convert any value of $Q$ to an actual fraction of satellite active BHs
above a given luminosity threshold. Values of $Q=1,2,4,6$ correspond to AGN
satellite fractions $f_\mathrm{sat}^\mathrm{AGN}= 0.10,0.18,0.31,0.40$,
respectively (see Table~\ref{tab:mock}). In Fig.~\ref{fig:mhalo_fsat} we
compare our predicted satellite fractions as a function of parent host DM halo
mass for different $Q$ values, with models of \citet{gatti16} and
\citet[][]{aird21} for AGNs with ${L_\mathrm{X}} > 10^{42}
\,\mathrm{erg}/\,\mathrm{s}$ and data from~\citet{pentericci13}
and~\cite{martini09}. Measurements of the AGN satellite fraction in groups and
clusters \citep{pentericci13, martini09} suggest small values of $Q \sim 1-2$,
as also assumed in the semi-empirical model presented in \citet[][]{aird21}. It
is worth noticing that independently of the parameter $Q$, $f_\mathrm{sat}$ is
by construction increasing as a function of the parent DM halo mass (see
Fig.~\ref{fig:mhalo_fsat}). In particular, higher $Q$ increases the AGN
satellite fraction in less massive parent halos. Our results show that the AGN
satellite fraction is thus an input parameter that controls the normalization
of the AGN large-scale bias as a function of stellar mass. Additional
measurements of $f_\mathrm{sat}$ in groups and clusters at this redshift would
help in putting independent constraints on $Q$.

\subsection{AGN HOD and the 1-halo term}

Finally, we estimate the projected two-point correlation function (2pcf) and
corresponding mean halo occupation distribution (HOD) of mock AGNs. In
particular, we measured the mock AGN 2pcf over the full scale range
$r_\mathrm{p} = 0.1-100 \,{h}^{-1}\mathrm{Mpc}$ including the small scales
within the 1-halo ($<1 {h}^{-1}\mathrm{Mpc}$) which is due to the correlation
of AGN within the same DM halo. This regime is especially sensitive to the
fraction of AGNs in satellite galaxies of galaxy groups and clusters. The mock
AGN HOD is calculated as the average number of mock AGNs in bins of DM halo
mass, separating the contribution of mock satellite and central AGNs.
Fig.~\ref{fig:gatti_wp} shows the 2pcf of mock AGNs with
$L_\mathrm{X}>10^{44}\,\mathrm{erg}/\mathrm{s}$ and ${N_\mathrm{H}} < 10^{22}
\,\mathrm{cm}^{-2}$ for $Q = 2,4$ and $6$, compared with the 2pcf estimated
\citet{richardson12}. The comparison with data suggests high values of $Q$ (=
4), with a corresponding satellite AGN fraction of
$f_\mathrm{sat}^\mathrm{AGN} 0.23$ (see Table~\ref{tab:mock}).

\begin{figure*}%
  \includegraphics[width=\figurewidthtwo]{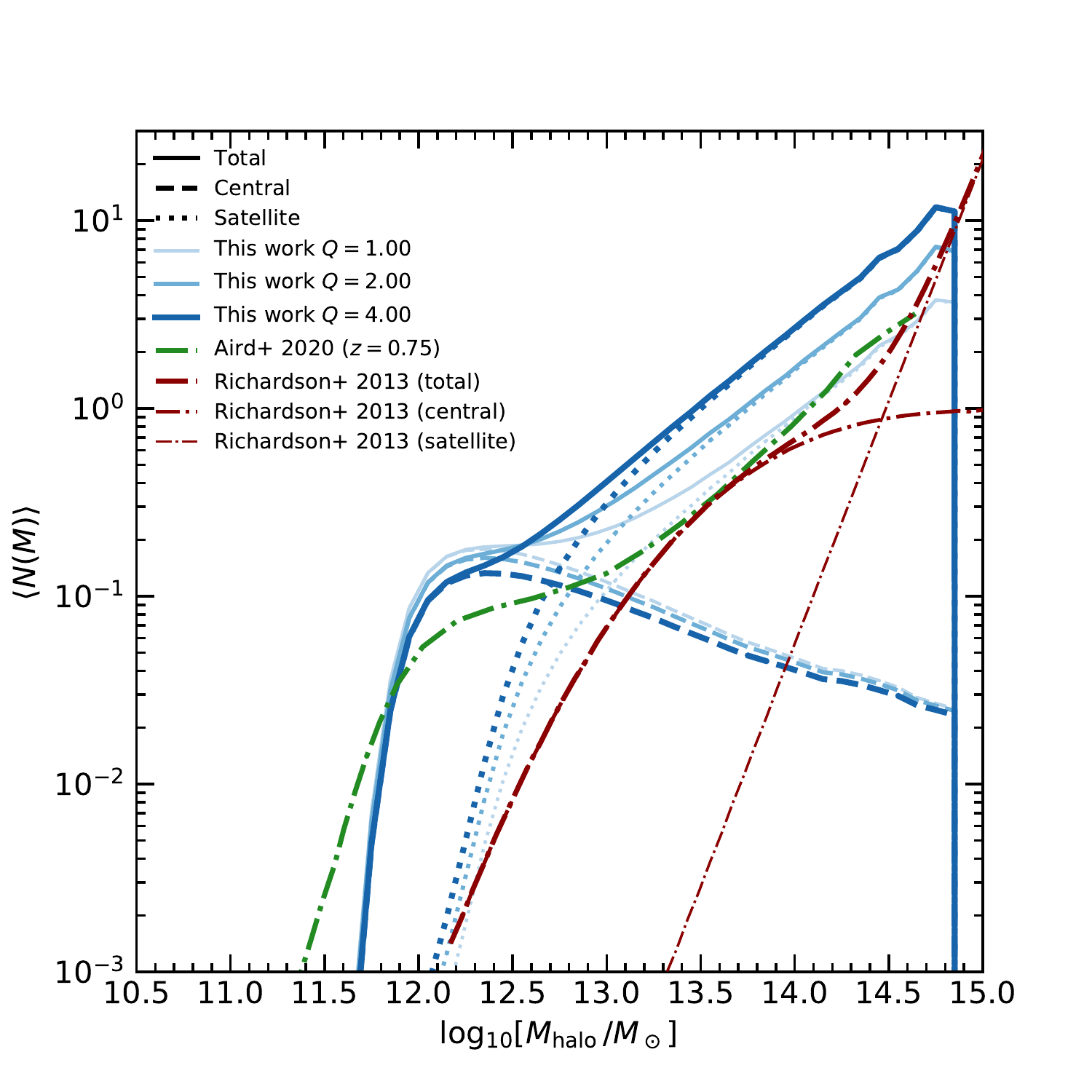}%
  \includegraphics[width=\figurewidthtwo]{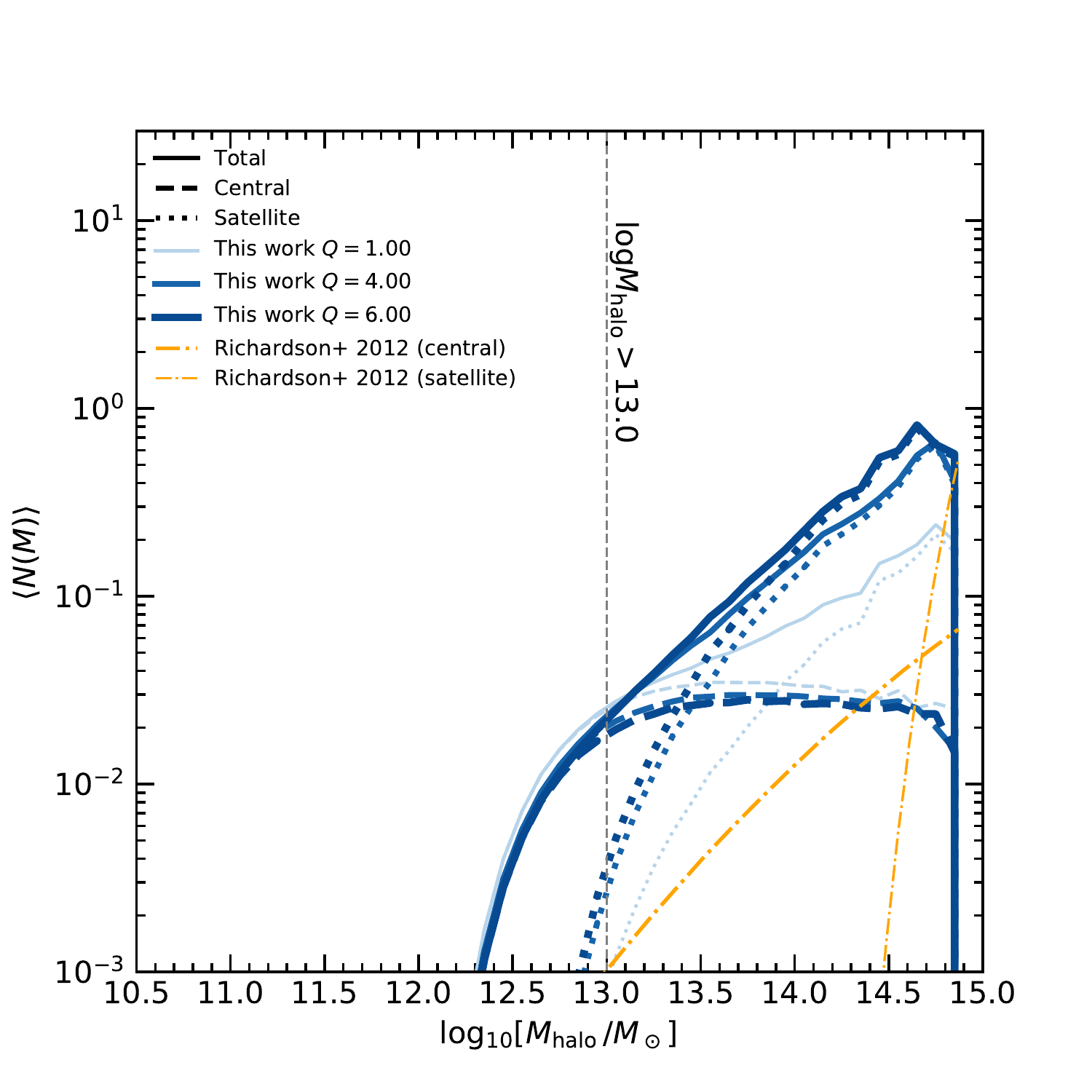}
  \caption{%
    Halo occupation distribution of mock AGNs derived by counting the average
    number of active BHs in bins of parent halo mass in this work (in blue) and
    in empirical mocks of \citet[][in green]{aird21}; and as estimated in
    \citet[][left panel, in red]{richardson13} and
    \citet[][right panel, in yellow]{richardson12}
    modelling the 2pcf of COSMOS AGN and SDSS quasars. Total, central, and
    satellite occupations are shown by solid, dashed, and dotted lines,
    respectively. Different linewidths correspond to different input $Q$ values
    as indicated by the legend. In addition, in the right panel we indicate as
    a vertical line the position of the model with the central DM halo mass
    cut, as shown in Fig.~\ref{fig:gatti_wp}.
  }%
  \label{fig:hod}
\end{figure*}

The corresponding mean HODs of mock AGNs and SDSS quasars are shown if
Fig.~\ref{fig:hod} (right panel). While the SDSS quasar HOD is characterized by
a significant central occupation only for DM halos with
$M_\mathrm{halo}>10^{13}\,M_{\odot}$ with a steep increase of the satellite
quasar HOD as a function of the hosting halo mass, the occupation of our
central mock AGNs is significant in halos with mass down to $M_\mathrm{halo}
\sim 10^{12.3} h^{-1}M_{\odot}$. Moreover, the satellite fraction derived for
SDSS quasars in \citet{richardson12} is $\sim 1000$ times smaller than the one
measured in our AGN mock catalog. This comparison shows that AGN small scale
clustering can be modelled by HODs characterized by different satellite AGN
fractions and minimum mass of the hosting halos.

A similar result is also shown in the left panel, for mock AGN with
$L_\mathrm{X}>10^{42}\,\mathrm{erg}/\mathrm{s}$, compared to the HOD derived in
\citet{richardson13} for X-ray selected COSMOS AGN at $z \sim 1.2$. Also for
moderate luminosity AGN, they found an upper limit of
$f_\mathrm{sat}^\mathrm{AGN} = 0.1$ and a significant central occupation for
central halos with mass $M_\mathrm{halo}>10^{12} h^{-1}M_{\odot}$, i.e.\ almost
3 times larger than what derived from our mock AGNs. For comparison, we also
show the AGN HOD from the semi-empirical model by \citet{aird21}, which is in
fair agreement with our predictions (for $Q = 1$), especially at large parent
halo masses. Their model also finds a significant AGN HOD for hosting halos
with mass down to $M_\mathrm{halo}>10^{11.5} h^{-1}M_{\odot}$.

\section{Discussion}%
\label{sec:discussion}

In this work we create realistic mock catalogs of active BHs and galaxies at $z
\sim 1.2$ based on semi-empirical models, to study the impact on the AGN
large-scale bias as a function of the host galaxy stellar mass and focus in
particular on the impact of (1) including a covariant scatter in the BH mass --
stellar mass relation at fixed halo mass; (2) changing the satellite AGN
fraction. We demonstrate that the AGN bias at fixed stellar mass is mainly
dependent on the scatter in the input scaling relations and on the relative
fraction of satellite and central active BHs.

\subsection{Covariant Scatter}

The scatter in the input BH mass -- stellar mass relation strongly affects the
AGN large-scale bias as a function of stellar mass. In particular we found that
the larger the scatter, the flatter is the resulting bias -- stellar mass
relation. This behaviour is expected as the end product of a covariant scatter
is to generate a larger scatter in the input In detail, the covariant scatter
model produces an AGN bias versus $M_\mathrm{star}$ up to twice as small at
$M_\mathrm{star} \sim 10^{11.5} M_{\odot}$ than the \emph{Reference} case. A
model with covariant scatter, i.e.\ with a (positive) correlation between BH
mass and galaxy mass at fixed halo mass, would then imply a bias as a function
of stellar mass substantially different between AGN and the overall population
of galaxies at the same stellar mass, at least for M$_\mathrm{star} > 10^{11}
M_{\odot}$, where unfortunately AGN bias estimates are not yet available and/or
still have large associated uncertainties
\citep{mountrichas19,viitanen19,allevato19}. At
$M_\mathrm{star}<10^{11}\,M_\odot$, instead, the difference in the large-scale
bias of normal and active galaxies at the same stellar mass is not significant,
as also found in \citet{mendez16} and \citet{powell18}.

Our predicted AGN large-scale bias is roughly consistent with the measurements
by \citep{mountrichas19,viitanen19}, which are however characterized by large
uncertainties, but it always falls short by a systematic $\sim 20\%$ in
reproducing the measured bias of CCL AGNs \citep{allevato19}, at least for
models with $Q=1$. The COSMOS field \citep[][]{scoville07} is known to be
characterized by large overdensities at $z < 1$ which could increase the
large-scale bias up to $\sim20-50\%$ \citep[][]{gilli09,mendez16,viitanen19}.
In fact, \citet{viitanen19} also estimate the AGN bias when removing from the
sample AGN in galaxy groups and measured a bias $\sim$0.4 dex smaller in the
lower stellar mass bin. In addition, the COSMOS field is affected by cosmic
variance due to the small volume of the survey which introduces an additional
$1 \sigma$ error of $\sim$0.1 dex to the bias measurements (see
Sec.\ref{sec:results} for more details). All these effects make it difficult to
constrain models with and without covariant scatter. In the near future,
precise clustering measurements of AGN that extend up to $M_\mathrm{star} >
10^{11} M_{\odot}$ will allow us to put constraints on these models.

The study of the degeneracy between AGN clustering strength and scatter in the
BH--host scaling relations has already been emphasized by a number of groups at
different redshifts
\citep[e.g.,][]{shen07,white08,wyithe09,bonoli10,shankar10,shankar10b}. One of
the main findings emphasised by these studies was the need for a small scatter
in the input scaling relation to boost the AGN clustering signal to match some
of the data. Our current results point to similar trends, whilst emphasizing
the vital role of additional parameters, such as the relative fraction of AGN
satellites, in amplifying the clustering signal.

\subsection{Satellite AGN fraction}

In this work we have highlighted the pivotal role of the $Q$ parameter, i.e.\
the ratio between active satellite and central BHs, in regulating the
normalization of the bias--stellar mass relation (e.g.,
Figure~\ref{fig:bias_q}). In addition, constraining the $Q$ parameter can shed
light on AGN triggering models. For example, a high relative fraction of
satellite AGN would be in line with the evidence put forward several times that
secular processes and bar instabilities, and not only mergers
\citep[e.g.,][]{hopkins08}, are efficient in triggering moderate-to-luminous
AGN
\citep[e.g.][]{georgakakis09,allevato11,cisternas11,schawinski11,rosario11}.

The value of $Q$ is yet not well constrained at $z>1$, whilst most studies
suggest that $Q\lesssim 2$ at $z \ge 1$. \citet[][]{allevato19} suggested that
the $z\sim 1$ large-scale bias of CCL Type 2 AGN as a function of the host
galaxy stellar mass can be reproduced assuming $Q \sim$ 2, which corresponds to
a satellite AGN fraction $f_\mathrm{AGN}^\mathrm{sat} \sim 0.15$, and applying
a cut in the AGN host halo mass of $M_\mathrm{halo} \gtrsim 10^{12}
h^{-1}M_{\odot}$. Estimates of the AGN content in groups of galaxies in the two
GOODS fields at $z \sim 1$ \citep{pentericci13} favour $Q \sim 1-2$. At smaller
redshifts, an AGN satellite fraction $f_\mathrm{AGN}^\mathrm{sat} \sim 0.18$
has been suggested by \citet[][]{leauthaud15} for COSMOS AGN at $z < 1$.
\citet[][]{allevato12} performed direct measurements of the HOD for COSMOS AGN
at $z < 0.2$ based on the mass function of galaxy groups hosting AGN, and found
that the duty cycle of satellite AGN is comparable or slightly larger than that
of central AGN, i.e. $Q \le 2$. \citet{georgakakis19} used semi-empirical
models to populate DM halos with AGN assuming no distinction between centrals
and satellites, i.e. $Q = 1$ and found a fair agreement, within the error
budget, between the observationally derived AGN HOD
\citep[e.g.][]{allevato12,miyaji11,shen13} and their AGN mock predictions.
\citet[][]{shankar20} also showed that the large-scale bias as a function of BH
mass of X-ray and optically selected AGN at $z \sim 0.25$ can be reproduced by
assuming $Q \le$ 2. Recently, \citet{allevato21} showed that the Q parameter
strongly affects the large-scale AGN bias as a function of stellar mass, BH
mass and luminosity, with $Q \sim 1-2$ more favoured by the data at $z \sim
0.1$.

We also investigated how the $Q$ parameter affects the AGN HOD and in turn the
1-halo term. The HOD approach has been used by different authors to interpret
AGN and quasar clustering measurements
\citep[e.g.][]{allevato11,miyaji11,richardson12,richardson13,kayo12}. In
particular, \citet{richardson12} and \citet{kayo12} performed clustering
measurements of optically luminous quasars for both small
($<1\,{h}^{-1}\mathrm{Mpc}$) and large physical scales and performed HOD
modelling to infer the relation between quasars and their host dark matter
halos (see Fig.~\ref{fig:hod}). \citet{richardson12} found at $z \sim 1.4$ a
small fraction of luminous SDSS quasars to be in satellites DM halos, with
$f_\mathrm{sat}^\mathrm{AGN}\sim 7.4 \times 10^{-4}$ and that the central
(satellite) occupation becomes significant only at masses above
$M_\mathrm{halo} \sim 10^{13} h^{-1}M_{\odot}$ ($\sim 10^{14}
h^{-1}M_{\odot}$). \citet{kayo12} also modelled the clustering of SDSS quasars
at $z \sim 1.4$ and reported a satellite fraction $\sim$100 times higher than
\citet{richardson12}, by using a different HOD parameterization. We also showed
that by increasing the input $Q$ parameter we were able to reproduce the full
2pcf of luminous SDSS quasars with resulting HODs very different from those
derived by \citet{richardson12}, further highlighting the degeneracies in HOD
modelling. We note that such degeneracies have already been reported from HOD
modeling for SDSS the quasar two-point cross-correlation function at
intermediate redshifts $0.3 < z < 0.9$ \citep{shen13}.

\subsection{Alternatives to high Q values}

The comparison of our mock predictions with the 2pcf at all scales of SDSS
quasars suggest high values of $Q$ ($\sim$4), which are in contrast with
previous measurements from HOD modelling in \citet{richardson12} and
\citet{kayo12} and previous findings at lower $z$
\citep[e.g.][]{leauthaud12,allevato12}. There could be various and concomitant
causes that could determine this offset. An explanation could be that the
MultiDark simulations used to create the AGN mock catalogs are characterized by
missing satellite halos possibly due to the low mass resolution
($1.5\times10^{9} h^{-1}M_{\odot}$) and/or stripped halos. Another possibility
is that luminous quasars do not reside in central DM halos with mass
$M_\mathrm{halo} < 10^{13} h^{-1}M_{\odot}$ as suggested by the HOD modelling
of \citet{richardson12}. In fact, as shown in Fig.~\ref{fig:gatti_wp}, the 2pcf
(including the 1-halo term) of SDSS quasars can be easily reproduced by mock
AGNs with $Q = 1$ and applying a cut in the minimum mass of central AGN hosting
halos at $M_\mathrm{halo} \sim 10^{13} h^{-1}M_{\odot}$. Similarly, one might
expect that the bias estimates as a function of stellar mass for COSMOS AGN
\citep{allevato19,viitanen19} can also be reproduced assuming smaller $Q$
($\sim$1) and a minimum mass in the central halos of $M_\mathrm{halo} \sim
10^{12} h^{-1}M_{\odot}$ \citep[following][]{richardson13}.%

These results suggest that the minimum central AGN hosting halo mass and
$f_\mathrm{sat}$ are degenerate. The smaller the AGN satellite fraction is, the
higher the mass needed for a central halo to host an AGN above a given
luminosity. Moreover, the smaller $f_\mathrm{sat}$, the steeper is the increase
of the satellite AGN occupation as a function of the parent halo mass. It is
certain that a fraction of quasars must be satellites to produce the
small-scale clustering of AGN\@. Currently available satellite AGN fraction
estimates in groups and clusters of galaxies presented by \citet{martini09} and
\citet{pentericci13} at $z \le 1$ suggest small values of $Q$. Additional
independent measurements of the AGN satellite fraction in groups and clusters
at $z \ge 1$ would help in breaking the degeneracy in these model parameters.

\section{Conclusions}%
\label{sec:conclusions}

Numerous degeneracies in the input parameters of cosmological models still
prevent solid progress on the open issue of the co-evolution of
supermassive black holes (BHs) and their host galaxies and dark matter halos.
Building on previous work from our group and by making use of advanced and
diverse semi-empirical routines, also inclusive of the covariance among
some input parameters, we here show that:
\begin{itemize}
  \item The overall shape and normalization of the large-scale bias as a
    function of AGN host galaxy stellar mass $b(M_\mathrm{star})$, is largely
    independent of the input stellar mass -- halo mass relation, duty cycle and
    Eddington ratio distribution, while it is mostly driven by the
    \emph{dispersion} in -- not so much by the shape of -- the input stellar
    mass--black hole mass relation.

  \item A model with covariant scatter, i.e.\ with a (positive) correlation
    between BH mass and galaxy mass at fixed halo mass, predicts an AGN bias
    almost independent of the stellar mass and substantially different from the
    bias of the underlying population of galaxies of the same stellar mass, at
    least in the range $M_\mathrm{star} > 10^{11} M_{\odot}$. Present AGN
    clustering estimates at $z \sim 1.2$ do not allow us to clearly distinguish
    between models with and without a covariant scatter.

  \item The other parameter controlling the normalization of the AGN bias
    $b(M_\mathrm{star})$ is $Q$, the relative fraction of AGN hosted in
    satellite and central BHs of a given mass. Increasing the probability of
    AGN to be hosted in satellites rather than in centrals of equal BH mass,
    naturally increases the large-scale clustering as the bias becomes more
    heavily weighted towards more massive host halos.

 \item The comparison with the large-scale bias of COSMOS AGN at $z\sim 1.2$
   and with the two-point correlation function of SDSS quasars at $z \sim 1.4$,
   suggests $Q \sim 4$ which corresponds to a relative fraction of AGN hosted
   in satellites $f_\mathrm{sat}^\mathrm{AGN} \sim 0.2-0.3$. However, the data
   are also reproduced by models that adopt $Q \le 2$, i.e.\ values more
   consistent with independent estimates at the AGN fraction at lower z, as
   long as a cut is applied in the minimum mass of central AGN hosting halos,
   as also suggested by the HOD modelling in some clustering studies. This
   result unveils a strong degeneracy between the AGN satellite fraction and
   the minimum halo mass hosting AGN above a given luminosity. Independent
   estimates of the fraction of active satellites in groups at $z \ge 1$ will
   help in breaking this degeneracy.
\end{itemize}

In the next years, current and imminent extragalactic surveys, such as Euclid,
eROSITA and LSST will precisely measure the clustering AGN at different masses
and redshifts allowing to set invaluable constraints on many important features
of AGN demography, such as limits on the covariance between AGN and galaxies,
on the minimum halo mass hosting AGN, on the relative fraction of AGN
satellites, and several others which will in turn provide essential constraints
on the still puzzling co-evolution of BHs and their host galaxies and dark
matter halos.

\section*{Acknowledgements}

The authors would like to thank the anonymous referee for their comments which
have signifcantly improved the manuscript. Additionally, the authors would like
to thank George Mountrichas for providing the bias measurement for the
\emph{XMM}-XXL AGN/VIPERS galaxies, and Hao Fu and Max Dickson for compiling
and sharing the stellar mass function estimates at $z \sim 1.2$.

AV is grateful to the Vilho, Yrjö and Kalle Väisälä Foundation of the Finnish
Academy of Science and Letters. VA acknowledges support from MUR-SNS-2019. FS
acknowledges partial support from a Leverhulme Trust Fellowship. This work was
supported by the Academy of Finland grant 295113.

The CosmoSim database used in this paper is a service by the Leibniz-Institute
for Astrophysics Potsdam (AIP).  The MultiDark database was developed in
cooperation with the Spanish MultiDark Consolider Project CSD2009-00064.

The authors gratefully acknowledge the Gauss Centre for Supercomputing e.V\@.
(\url{www.gauss-centre.eu}) and the Partnership for Advanced Supercomputing in
Europe (PRACE, \url{www.prace-ri.eu}) for funding the MultiDark simulation
project by providing computing time on the GCS Supercomputer SuperMUC at
Leibniz Supercomputing Centre (LRZ, \url{www.lrz.de}). The Bolshoi simulations
have been performed within the Bolshoi project of the University of California
High-Performance AstroComputing Center (UC-HiPACC) and were run at the NASA
Ames Research Center.

This research has made use of NASA’s Astrophysics Data System.

\section*{Data Availability}

The MultiDark ROCKSTAR halo catalogues are available in the CosmoSim database
at \url{https://www.cosmosim.org/}. Other data underlying this article will
be shared on reasonable request to the corresponding author.

\bibliographystyle{mnras}

\begin{thebibliography}{}
\makeatletter
\relax
\def\mn@urlcharsother{\let\do\@makeother \do\$\do\&\do\#\do\^\do\_\do\%\do\~}
\def\mn@doi{\begingroup\mn@urlcharsother \@ifnextchar [ {\mn@doi@}
  {\mn@doi@[]}}
\def\mn@doi@[#1]#2{\def\@tempa{#1}\ifx\@tempa\@empty \href
  {http://dx.doi.org/#2} {doi:#2}\else \href {http://dx.doi.org/#2} {#1}\fi
  \endgroup}
\def\mn@eprint#1#2{\mn@eprint@#1:#2::\@nil}
\def\mn@eprint@arXiv#1{\href {http://arxiv.org/abs/#1} {{\tt arXiv:#1}}}
\def\mn@eprint@dblp#1{\href {http://dblp.uni-trier.de/rec/bibtex/#1.xml}
  {dblp:#1}}
\def\mn@eprint@#1:#2:#3:#4\@nil{\def\@tempa {#1}\def\@tempb {#2}\def\@tempc
  {#3}\ifx \@tempc \@empty \let \@tempc \@tempb \let \@tempb \@tempa \fi \ifx
  \@tempb \@empty \def\@tempb {arXiv}\fi \@ifundefined
  {mn@eprint@\@tempb}{\@tempb:\@tempc}{\expandafter \expandafter \csname
  mn@eprint@\@tempb\endcsname \expandafter{\@tempc}}}

\bibitem[\protect\citeauthoryear{{Aird} \& {Coil}}{{Aird} \&
  {Coil}}{2021}]{aird21}
{Aird} J.,  {Coil} A.~L.,  2021, \mn@doi [\mnras] {10.1093/mnras/stab312},
  \href {https://ui.adsabs.harvard.edu/abs/2021MNRAS.502.5962A} {502, 5962}

\bibitem[\protect\citeauthoryear{{Aird}, {Coil}  \& {Georgakakis}}{{Aird}
  et~al.}{2017}]{aird17}
{Aird} J.,  {Coil} A.~L.,   {Georgakakis} A.,  2017, \mn@doi [\mnras]
  {10.1093/mnras/stw2932}, \href
  {https://ui.adsabs.harvard.edu/abs/2017MNRAS.465.3390A} {465, 3390}

\bibitem[\protect\citeauthoryear{{Allevato} et~al.,}{{Allevato}
  et~al.}{2011}]{allevato11}
{Allevato} V.,  et~al., 2011, \mn@doi [\apj] {10.1088/0004-637X/736/2/99},
  \href {http://adsabs.harvard.edu/abs/2011ApJ...736...99A} {736, 99}

\bibitem[\protect\citeauthoryear{{Allevato} et~al.,}{{Allevato}
  et~al.}{2012}]{allevato12}
{Allevato} V.,  et~al., 2012, \mn@doi [\apj] {10.1088/0004-637X/758/1/47},
  \href {http://adsabs.harvard.edu/abs/2012ApJ...758...47A} {758, 47}

\bibitem[\protect\citeauthoryear{{Allevato} et~al.,}{{Allevato}
  et~al.}{2019}]{allevato19}
{Allevato} V.,  et~al., 2019, \mn@doi [\aap] {10.1051/0004-6361/201936191},
  \href {https://ui.adsabs.harvard.edu/abs/2019A&A...632A..88A} {632, A88}

\bibitem[\protect\citeauthoryear{{Allevato}, {Shankar}, {Marsden}, {Rasulov},
  {Viitanen}, {Georgakakis}, {Ferrara}  \& {Finoguenov}}{{Allevato}
  et~al.}{2021}]{allevato21}
{Allevato} V.,  {Shankar} F.,  {Marsden} C.,  {Rasulov} U.,  {Viitanen} A.,
  {Georgakakis} A.,  {Ferrara} A.,   {Finoguenov} A.,  2021, arXiv e-prints,
  \href {https://ui.adsabs.harvard.edu/abs/2021arXiv210502883A} {p.
  arXiv:2105.02883}

\bibitem[\protect\citeauthoryear{{Alonso}}{{Alonso}}{2012}]{alonso12}
{Alonso} D.,  2012, arXiv e-prints, \href
  {https://ui.adsabs.harvard.edu/abs/2012arXiv1210.1833A} {p. arXiv:1210.1833}

\bibitem[\protect\citeauthoryear{{Aversa}, {Lapi}, {de Zotti}, {Shankar}  \&
  {Danese}}{{Aversa} et~al.}{2015}]{aversa15}
{Aversa} R.,  {Lapi} A.,  {de Zotti} G.,  {Shankar} F.,   {Danese} L.,  2015,
  \mn@doi [\apj] {10.1088/0004-637X/810/1/74}, \href
  {https://ui.adsabs.harvard.edu/abs/2015ApJ...810...74A} {810, 74}

\bibitem[\protect\citeauthoryear{{Bandara}, {Crampton}  \& {Simard}}{{Bandara}
  et~al.}{2009}]{bandara09}
{Bandara} K.,  {Crampton} D.,   {Simard} L.,  2009, \mn@doi [\apj]
  {10.1088/0004-637X/704/2/1135}, \href
  {https://ui.adsabs.harvard.edu/abs/2009ApJ...704.1135B} {704, 1135}

\bibitem[\protect\citeauthoryear{{Behroozi}, {Wechsler}  \& {Wu}}{{Behroozi}
  et~al.}{2013}]{behroozi13a}
{Behroozi} P.~S.,  {Wechsler} R.~H.,   {Wu} H.-Y.,  2013, \mn@doi [\apj]
  {10.1088/0004-637X/762/2/109}, \href
  {http://adsabs.harvard.edu/abs/2013ApJ...762..109B} {762, 109}

\bibitem[\protect\citeauthoryear{{Behroozi}, {Wechsler}, {Hearin}  \&
  {Conroy}}{{Behroozi} et~al.}{2019}]{behroozi19}
{Behroozi} P.,  {Wechsler} R.~H.,  {Hearin} A.~P.,   {Conroy} C.,  2019,
  \mn@doi [\mnras] {10.1093/mnras/stz1182}, \href
  {https://ui.adsabs.harvard.edu/abs/2019MNRAS.488.3143B} {488, 3143}

\bibitem[\protect\citeauthoryear{{Bernardi}}{{Bernardi}}{2007}]{bernardi07}
{Bernardi} M.,  2007, \mn@doi [\aj] {10.1086/512611}, \href
  {https://ui.adsabs.harvard.edu/abs/2007AJ....133.1954B} {133, 1954}

\bibitem[\protect\citeauthoryear{{Bongiorno} et~al.,}{{Bongiorno}
  et~al.}{2012}]{bongiorno12}
{Bongiorno} A.,  et~al., 2012, \mn@doi [\mnras]
  {10.1111/j.1365-2966.2012.22089.x}, \href
  {http://adsabs.harvard.edu/abs/2012MNRAS.427.3103B} {427, 3103}

\bibitem[\protect\citeauthoryear{{Bongiorno} et~al.,}{{Bongiorno}
  et~al.}{2016}]{bongiorno16}
{Bongiorno} A.,  et~al., 2016, \mn@doi [\aap] {10.1051/0004-6361/201527436},
  \href {http://adsabs.harvard.edu/abs/2016A%26A...588A..78B} {588, A78}

\bibitem[\protect\citeauthoryear{{Bonoli}, {Shankar}, {White}, {Springel}  \&
  {Wyithe}}{{Bonoli} et~al.}{2010}]{bonoli10}
{Bonoli} S.,  {Shankar} F.,  {White} S. D.~M.,  {Springel} V.,   {Wyithe} J.
  S.~B.,  2010, \mn@doi [\mnras] {10.1111/j.1365-2966.2010.16285.x}, \href
  {https://ui.adsabs.harvard.edu/abs/2010MNRAS.404..399B} {404, 399}

\bibitem[\protect\citeauthoryear{{Cisternas} et~al.,}{{Cisternas}
  et~al.}{2011a}]{cisternas11}
{Cisternas} M.,  et~al., 2011a, \mn@doi [\apj] {10.1088/0004-637X/726/2/57},
  \href {https://ui.adsabs.harvard.edu/abs/2011ApJ...726...57C} {726, 57}

\bibitem[\protect\citeauthoryear{{Cisternas} et~al.,}{{Cisternas}
  et~al.}{2011b}]{cisternas11b}
{Cisternas} M.,  et~al., 2011b, \mn@doi [\apjl] {10.1088/2041-8205/741/1/L11},
  \href {https://ui.adsabs.harvard.edu/abs/2011ApJ...741L..11C} {741, L11}

\bibitem[\protect\citeauthoryear{{Coil} et~al.,}{{Coil} et~al.}{2009}]{coil09}
{Coil} A.~L.,  et~al., 2009, \mn@doi [\apj] {10.1088/0004-637X/701/2/1484},
  \href {http://adsabs.harvard.edu/abs/2009ApJ...701.1484C} {701, 1484}

\bibitem[\protect\citeauthoryear{{Davidzon} et~al.,}{{Davidzon}
  et~al.}{2017}]{davidzon17}
{Davidzon} I.,  et~al., 2017, \mn@doi [\aap] {10.1051/0004-6361/201730419},
  \href {https://ui.adsabs.harvard.edu/abs/2017A&A...605A..70D} {605, A70}

\bibitem[\protect\citeauthoryear{{Davis} \& {Peebles}}{{Davis} \&
  {Peebles}}{1983}]{davis_peebles83}
{Davis} M.,  {Peebles} P.~J.~E.,  1983, \mn@doi [\apj] {10.1086/160884}, \href
  {http://adsabs.harvard.edu/abs/1983ApJ...267..465D} {267, 465}

\bibitem[\protect\citeauthoryear{{Davis}, {Graham}  \& {Cameron}}{{Davis}
  et~al.}{2019}]{davis19}
{Davis} B.~L.,  {Graham} A.~W.,   {Cameron} E.,  2019, \mn@doi [\apj]
  {10.3847/1538-4357/aaf3b8}, \href
  {https://ui.adsabs.harvard.edu/abs/2019ApJ...873...85D} {873, 85}

\bibitem[\protect\citeauthoryear{{Eisenstein} \& {Hu}}{{Eisenstein} \&
  {Hu}}{1998}]{eisenstein_hu98}
{Eisenstein} D.~J.,  {Hu} W.,  1998, \mn@doi [\apj] {10.1086/305424}, \href
  {http://adsabs.harvard.edu/abs/1998ApJ...496..605E} {496, 605}

\bibitem[\protect\citeauthoryear{{Fanidakis} et~al.,}{{Fanidakis}
  et~al.}{2012}]{fanidakis12}
{Fanidakis} N.,  et~al., 2012, \mn@doi [\mnras]
  {10.1111/j.1365-2966.2011.19931.x}, \href
  {http://adsabs.harvard.edu/abs/2012MNRAS.419.2797F} {419, 2797}

\bibitem[\protect\citeauthoryear{{Farahi} et~al.,}{{Farahi}
  et~al.}{2019}]{farahi19}
{Farahi} A.,  et~al., 2019, \mn@doi [Nature Communications]
  {10.1038/s41467-019-10471-y}, \href
  {https://ui.adsabs.harvard.edu/abs/2019NatCo..10.2504F} {10, 2504}

\bibitem[\protect\citeauthoryear{{Ferrarese}}{{Ferrarese}}{2002}]{ferrarese02}
{Ferrarese} L.,  2002, \mn@doi [\apj] {10.1086/342308}, \href
  {https://ui.adsabs.harvard.edu/abs/2002ApJ...578...90F} {578, 90}

\bibitem[\protect\citeauthoryear{{Ferrarese} \& {Ford}}{{Ferrarese} \&
  {Ford}}{2005}]{ferrarese_ford05}
{Ferrarese} L.,  {Ford} H.,  2005, \mn@doi [\ssr] {10.1007/s11214-005-3947-6},
  \href {http://adsabs.harvard.edu/abs/2005SSRv..116..523F} {116, 523}

\bibitem[\protect\citeauthoryear{{Gatti}, {Shankar}, {Bouillot}, {Menci},
  {Lamastra}, {Hirschmann}  \& {Fiore}}{{Gatti} et~al.}{2016}]{gatti16}
{Gatti} M.,  {Shankar} F.,  {Bouillot} V.,  {Menci} N.,  {Lamastra} A.,
  {Hirschmann} M.,   {Fiore} F.,  2016, \mn@doi [\mnras]
  {10.1093/mnras/stv2754}, \href
  {http://adsabs.harvard.edu/abs/2016MNRAS.456.1073G} {456, 1073}

\bibitem[\protect\citeauthoryear{{Georgakakis} et~al.,}{{Georgakakis}
  et~al.}{2009}]{georgakakis09}
{Georgakakis} A.,  et~al., 2009, \mn@doi [\mnras]
  {10.1111/j.1365-2966.2009.14951.x}, \href
  {https://ui.adsabs.harvard.edu/abs/2009MNRAS.397..623G} {397, 623}

\bibitem[\protect\citeauthoryear{{Georgakakis} et~al.,}{{Georgakakis}
  et~al.}{2014}]{georgakakis14}
{Georgakakis} A.,  et~al., 2014, \mn@doi [\mnras] {10.1093/mnras/stu1326},
  \href {http://adsabs.harvard.edu/abs/2014MNRAS.443.3327G} {443, 3327}

\bibitem[\protect\citeauthoryear{{Georgakakis}, {Aird}, {Schulze}, {Dwelly},
  {Salvato}, {Nandra}, {Merloni}  \& {Schneider}}{{Georgakakis}
  et~al.}{2017}]{georgakakis17}
{Georgakakis} A.,  {Aird} J.,  {Schulze} A.,  {Dwelly} T.,  {Salvato} M.,
  {Nandra} K.,  {Merloni} A.,   {Schneider} D.~P.,  2017, \mn@doi [\mnras]
  {10.1093/mnras/stx1602}, \href
  {http://adsabs.harvard.edu/abs/2017MNRAS.471.1976G} {471, 1976}

\bibitem[\protect\citeauthoryear{{Georgakakis}, {Comparat}, {Merloni},
  {Ciesla}, {Aird}  \& {Finoguenov}}{{Georgakakis}
  et~al.}{2019}]{georgakakis19}
{Georgakakis} A.,  {Comparat} J.,  {Merloni} A.,  {Ciesla} L.,  {Aird} J.,
  {Finoguenov} A.,  2019, \mn@doi [\mnras] {10.1093/mnras/sty3454}, \href
  {https://ui.adsabs.harvard.edu/abs/2019MNRAS.487..275G} {487, 275}

\bibitem[\protect\citeauthoryear{{Gilli} et~al.,}{{Gilli}
  et~al.}{2009}]{gilli09}
{Gilli} R.,  et~al., 2009, \mn@doi [\aap] {10.1051/0004-6361:200810821}, \href
  {http://adsabs.harvard.edu/abs/2009A%26A...494...33G} {494, 33}

\bibitem[\protect\citeauthoryear{{Graham} \& {Scott}}{{Graham} \&
  {Scott}}{2015}]{graham15}
{Graham} A.~W.,  {Scott} N.,  2015, \mn@doi [\apj]
  {10.1088/0004-637X/798/1/54}, \href
  {https://ui.adsabs.harvard.edu/abs/2015ApJ...798...54G} {798, 54}

\bibitem[\protect\citeauthoryear{{Granato}, {De Zotti}, {Silva}, {Bressan}  \&
  {Danese}}{{Granato} et~al.}{2004}]{granato04}
{Granato} G.~L.,  {De Zotti} G.,  {Silva} L.,  {Bressan} A.,   {Danese} L.,
  2004, \mn@doi [\apj] {10.1086/379875}, \href
  {https://ui.adsabs.harvard.edu/abs/2004ApJ...600..580G} {600, 580}

\bibitem[\protect\citeauthoryear{{Grylls}, {Shankar}, {Zanisi}  \&
  {Bernardi}}{{Grylls} et~al.}{2019}]{grylls19}
{Grylls} P.~J.,  {Shankar} F.,  {Zanisi} L.,   {Bernardi} M.,  2019, \mn@doi
  [\mnras] {10.1093/mnras/sty3281}, \href
  {https://ui.adsabs.harvard.edu/abs/2019MNRAS.483.2506G} {483, 2506}

\bibitem[\protect\citeauthoryear{{Hickox} et~al.,}{{Hickox}
  et~al.}{2009}]{hickox09a}
{Hickox} R.~C.,  et~al., 2009, \mn@doi [\apj] {10.1088/0004-637X/696/1/891},
  \href {http://adsabs.harvard.edu/abs/2009ApJ...696..891H} {696, 891}

\bibitem[\protect\citeauthoryear{{Hirschmann}, {Khochfar}, {Burkert}, {Naab},
  {Genel}  \& {Somerville}}{{Hirschmann} et~al.}{2010}]{hirschmann10}
{Hirschmann} M.,  {Khochfar} S.,  {Burkert} A.,  {Naab} T.,  {Genel} S.,
  {Somerville} R.~S.,  2010, \mn@doi [\mnras]
  {10.1111/j.1365-2966.2010.17006.x}, \href
  {https://ui.adsabs.harvard.edu/abs/2010MNRAS.407.1016H} {407, 1016}

\bibitem[\protect\citeauthoryear{{Hopkins}, {Hernquist}, {Cox}  \&
  {Kere{\v{s}}}}{{Hopkins} et~al.}{2008}]{hopkins08}
{Hopkins} P.~F.,  {Hernquist} L.,  {Cox} T.~J.,   {Kere{\v{s}}} D.,  2008,
  \mn@doi [\apjs] {10.1086/524362}, \href
  {https://ui.adsabs.harvard.edu/abs/2008ApJS..175..356H} {175, 356}

\bibitem[\protect\citeauthoryear{{Jahnke} \& {Macci{\`o}}}{{Jahnke} \&
  {Macci{\`o}}}{2011}]{jahnke11}
{Jahnke} K.,  {Macci{\`o}} A.~V.,  2011, \mn@doi [\apj]
  {10.1088/0004-637X/734/2/92}, \href
  {https://ui.adsabs.harvard.edu/abs/2011ApJ...734...92J} {734, 92}

\bibitem[\protect\citeauthoryear{{Jullo} et~al.,}{{Jullo}
  et~al.}{2012}]{jullo12}
{Jullo} E.,  et~al., 2012, \mn@doi [\apj] {10.1088/0004-637X/750/1/37}, \href
  {https://ui.adsabs.harvard.edu/abs/2012ApJ...750...37J} {750, 37}

\bibitem[\protect\citeauthoryear{{Kayo} \& {Oguri}}{{Kayo} \&
  {Oguri}}{2012}]{kayo12}
{Kayo} I.,  {Oguri} M.,  2012, \mn@doi [\mnras]
  {10.1111/j.1365-2966.2012.21321.x}, \href
  {https://ui.adsabs.harvard.edu/abs/2012MNRAS.424.1363K} {424, 1363}

\bibitem[\protect\citeauthoryear{{Kelly}}{{Kelly}}{2007}]{kelly07}
{Kelly} B.~C.,  2007, \mn@doi [\apj] {10.1086/519947}, \href
  {https://ui.adsabs.harvard.edu/abs/2007ApJ...665.1489K} {665, 1489}

\bibitem[\protect\citeauthoryear{{Klypin}, {Yepes}, {Gottl{\"o}ber}, {Prada}
  \& {He{\ss}}}{{Klypin} et~al.}{2016}]{klypin16}
{Klypin} A.,  {Yepes} G.,  {Gottl{\"o}ber} S.,  {Prada} F.,   {He{\ss}} S.,
  2016, \mn@doi [\mnras] {10.1093/mnras/stw248}, \href
  {https://ui.adsabs.harvard.edu/abs/2016MNRAS.457.4340K} {457, 4340}

\bibitem[\protect\citeauthoryear{{Kormendy} \& {Ho}}{{Kormendy} \&
  {Ho}}{2013}]{kormendy_ho13}
{Kormendy} J.,  {Ho} L.~C.,  2013, \mn@doi [\araa]
  {10.1146/annurev-astro-082708-101811}, \href
  {http://adsabs.harvard.edu/abs/2013ARA%26A..51..511K} {51, 511}

\bibitem[\protect\citeauthoryear{{Krumpe}, {Miyaji}  \& {Coil}}{{Krumpe}
  et~al.}{2010}]{krumpe10}
{Krumpe} M.,  {Miyaji} T.,   {Coil} A.~L.,  2010, \mn@doi [\apj]
  {10.1088/0004-637X/713/1/558}, \href
  {http://adsabs.harvard.edu/abs/2010ApJ...713..558K} {713, 558}

\bibitem[\protect\citeauthoryear{{Krumpe}, {Miyaji}, {Coil}  \&
  {Aceves}}{{Krumpe} et~al.}{2012}]{krumpe12}
{Krumpe} M.,  {Miyaji} T.,  {Coil} A.~L.,   {Aceves} H.,  2012, \mn@doi [\apj]
  {10.1088/0004-637X/746/1/1}, \href
  {http://adsabs.harvard.edu/abs/2012ApJ...746....1K} {746, 1}

\bibitem[\protect\citeauthoryear{{Leauthaud} et~al.,}{{Leauthaud}
  et~al.}{2012}]{leauthaud12}
{Leauthaud} A.,  et~al., 2012, \mn@doi [\apj] {10.1088/0004-637X/744/2/159},
  \href {https://ui.adsabs.harvard.edu/abs/2012ApJ...744..159L} {744, 159}

\bibitem[\protect\citeauthoryear{{Leauthaud} et~al.,}{{Leauthaud}
  et~al.}{2015}]{leauthaud15}
{Leauthaud} A.,  et~al., 2015, \mn@doi [\mnras] {10.1093/mnras/stu2210}, \href
  {http://adsabs.harvard.edu/abs/2015MNRAS.446.1874L} {446, 1874}

\bibitem[\protect\citeauthoryear{{Madau} \& {Dickinson}}{{Madau} \&
  {Dickinson}}{2014}]{madau_dickinson14}
{Madau} P.,  {Dickinson} M.,  2014, \mn@doi [\araa]
  {10.1146/annurev-astro-081811-125615}, \href
  {https://ui.adsabs.harvard.edu/abs/2014ARA&A..52..415M} {52, 415}

\bibitem[\protect\citeauthoryear{{Marconi}, {Risaliti}, {Gilli}, {Hunt},
  {Maiolino}  \& {Salvati}}{{Marconi} et~al.}{2004}]{marconi04}
{Marconi} A.,  {Risaliti} G.,  {Gilli} R.,  {Hunt} L.~K.,  {Maiolino} R.,
  {Salvati} M.,  2004, \mn@doi [\mnras] {10.1111/j.1365-2966.2004.07765.x},
  \href {http://adsabs.harvard.edu/abs/2004MNRAS.351..169M} {351, 169}

\bibitem[\protect\citeauthoryear{{Marsden}, {Shankar}, {Ginolfi}  \&
  {Zubovas}}{{Marsden} et~al.}{2020}]{marsden20}
{Marsden} C.,  {Shankar} F.,  {Ginolfi} M.,   {Zubovas} K.,  2020, \mn@doi
  [Frontiers in Physics] {10.3389/fphy.2020.00061}, \href
  {https://ui.adsabs.harvard.edu/abs/2020FrP.....8...61M} {8, 61}

\bibitem[\protect\citeauthoryear{{Martini}, {Sivakoff}  \&
  {Mulchaey}}{{Martini} et~al.}{2009}]{martini09}
{Martini} P.,  {Sivakoff} G.~R.,   {Mulchaey} J.~S.,  2009, \mn@doi [\apj]
  {10.1088/0004-637X/701/1/66}, \href
  {https://ui.adsabs.harvard.edu/abs/2009ApJ...701...66M} {701, 66}

\bibitem[\protect\citeauthoryear{{Marulli} et~al.,}{{Marulli}
  et~al.}{2013}]{marulli13}
{Marulli} F.,  et~al., 2013, \mn@doi [\aap] {10.1051/0004-6361/201321476},
  \href {http://adsabs.harvard.edu/abs/2013A%26A...557A..17M} {557, A17}

\bibitem[\protect\citeauthoryear{{Mendez} et~al.,}{{Mendez}
  et~al.}{2016}]{mendez16}
{Mendez} A.~J.,  et~al., 2016, \mn@doi [\apj] {10.3847/0004-637X/821/1/55},
  \href {http://adsabs.harvard.edu/abs/2016ApJ...821...55M} {821, 55}

\bibitem[\protect\citeauthoryear{{Miyaji}, {Krumpe}, {Coil}  \&
  {Aceves}}{{Miyaji} et~al.}{2011}]{miyaji11}
{Miyaji} T.,  {Krumpe} M.,  {Coil} A.~L.,   {Aceves} H.,  2011, \mn@doi [\apj]
  {10.1088/0004-637X/726/2/83}, \href
  {http://adsabs.harvard.edu/abs/2011ApJ...726...83M} {726, 83}

\bibitem[\protect\citeauthoryear{{Miyaji} et~al.,}{{Miyaji}
  et~al.}{2015}]{miyaji15}
{Miyaji} T.,  et~al., 2015, \mn@doi [\apj] {10.1088/0004-637X/804/2/104}, \href
  {https://ui.adsabs.harvard.edu/abs/2015ApJ...804..104M} {804, 104}

\bibitem[\protect\citeauthoryear{{Moster}, {Somerville}, {Maulbetsch}, {van den
  Bosch}, {Macci{\`o}}, {Naab}  \& {Oser}}{{Moster} et~al.}{2010}]{moster10}
{Moster} B.~P.,  {Somerville} R.~S.,  {Maulbetsch} C.,  {van den Bosch} F.~C.,
  {Macci{\`o}} A.~V.,  {Naab} T.,   {Oser} L.,  2010, \mn@doi [\apj]
  {10.1088/0004-637X/710/2/903}, \href
  {http://adsabs.harvard.edu/abs/2010ApJ...710..903M} {710, 903}

\bibitem[\protect\citeauthoryear{{Mountrichas} \& {Georgakakis}}{{Mountrichas}
  \& {Georgakakis}}{2012}]{mountrichas_georgakakis12}
{Mountrichas} G.,  {Georgakakis} A.,  2012, \mn@doi [\mnras]
  {10.1111/j.1365-2966.2011.20059.x}, \href
  {http://adsabs.harvard.edu/abs/2012MNRAS.420..514M} {420, 514}

\bibitem[\protect\citeauthoryear{{Mountrichas}, {Georgakakis}  \&
  {Georgantopoulos}}{{Mountrichas} et~al.}{2019}]{mountrichas19}
{Mountrichas} G.,  {Georgakakis} A.,   {Georgantopoulos} I.,  2019, \mn@doi
  [\mnras] {10.1093/mnras/sty3140}, \href
  {http://adsabs.harvard.edu/abs/2019MNRAS.483.1374M} {483, 1374}

\bibitem[\protect\citeauthoryear{{Peebles}}{{Peebles}}{1980}]{peebles80}
{Peebles} P.~J.~E.,  1980, {The large-scale structure of the universe}.
Princeton University Press

\bibitem[\protect\citeauthoryear{{Pentericci} et~al.,}{{Pentericci}
  et~al.}{2013}]{pentericci13}
{Pentericci} L.,  et~al., 2013, \mn@doi [\aap] {10.1051/0004-6361/201219759},
  \href {https://ui.adsabs.harvard.edu/abs/2013A&A...552A.111P} {552, A111}

\bibitem[\protect\citeauthoryear{{Powell} et~al.,}{{Powell}
  et~al.}{2018}]{powell18}
{Powell} M.~C.,  et~al., 2018, \mn@doi [\apj] {10.3847/1538-4357/aabd7f}, \href
  {http://adsabs.harvard.edu/abs/2018ApJ...858..110P} {858, 110}

\bibitem[\protect\citeauthoryear{{Powell}, {Urry}, {Cappelluti}, {Johnson},
  {LaMassa}, {Ananna}  \& {Kollmann}}{{Powell} et~al.}{2020}]{powell20}
{Powell} M.~C.,  {Urry} C.~M.,  {Cappelluti} N.,  {Johnson} J.~T.,  {LaMassa}
  S.~M.,  {Ananna} T.~T.,   {Kollmann} K.~E.,  2020, \mn@doi [\apj]
  {10.3847/1538-4357/ab6e65}, \href
  {https://ui.adsabs.harvard.edu/abs/2020ApJ...891...41P} {891, 41}

\bibitem[\protect\citeauthoryear{{Reines} \& {Volonteri}}{{Reines} \&
  {Volonteri}}{2015}]{reines15}
{Reines} A.~E.,  {Volonteri} M.,  2015, \mn@doi [\apj]
  {10.1088/0004-637X/813/2/82}, \href
  {https://ui.adsabs.harvard.edu/abs/2015ApJ...813...82R} {813, 82}

\bibitem[\protect\citeauthoryear{{Ricci}, {Marchesi}, {Shankar}, {La Franca}
  \& {Civano}}{{Ricci} et~al.}{2017}]{ricci17}
{Ricci} F.,  {Marchesi} S.,  {Shankar} F.,  {La Franca} F.,   {Civano} F.,
  2017, \mn@doi [\mnras] {10.1093/mnras/stw2909}, \href
  {https://ui.adsabs.harvard.edu/abs/2017MNRAS.465.1915R} {465, 1915}

\bibitem[\protect\citeauthoryear{{Richardson}, {Zheng}, {Chatterjee}, {Nagai}
  \& {Shen}}{{Richardson} et~al.}{2012}]{richardson12}
{Richardson} J.,  {Zheng} Z.,  {Chatterjee} S.,  {Nagai} D.,   {Shen} Y.,
  2012, \mn@doi [\apj] {10.1088/0004-637X/755/1/30}, \href
  {http://adsabs.harvard.edu/abs/2012ApJ...755...30R} {755, 30}

\bibitem[\protect\citeauthoryear{{Richardson}, {Chatterjee}, {Zheng}, {Myers}
  \& {Hickox}}{{Richardson} et~al.}{2013}]{richardson13}
{Richardson} J.,  {Chatterjee} S.,  {Zheng} Z.,  {Myers} A.~D.,   {Hickox} R.,
  2013, \mn@doi [\apj] {10.1088/0004-637X/774/2/143}, \href
  {https://ui.adsabs.harvard.edu/abs/2013ApJ...774..143R} {774, 143}

\bibitem[\protect\citeauthoryear{{Riebe} et~al.,}{{Riebe}
  et~al.}{2013}]{riebe13}
{Riebe} K.,  et~al., 2013, \mn@doi [Astronomische Nachrichten]
  {10.1002/asna.201211900}, \href
  {http://adsabs.harvard.edu/abs/2013AN....334..691R} {334, 691}

\bibitem[\protect\citeauthoryear{{Rodr{\'\i}guez-Puebla}, {Avila-Reese},
  {Yang}, {Foucaud}, {Drory}  \& {Jing}}{{Rodr{\'\i}guez-Puebla}
  et~al.}{2015}]{rodriguez-puebla15}
{Rodr{\'\i}guez-Puebla} A.,  {Avila-Reese} V.,  {Yang} X.,  {Foucaud} S.,
  {Drory} N.,   {Jing} Y.~P.,  2015, \mn@doi [\apj]
  {10.1088/0004-637X/799/2/130}, \href
  {https://ui.adsabs.harvard.edu/abs/2015ApJ...799..130R} {799, 130}

\bibitem[\protect\citeauthoryear{{Rosario}, {McGurk}, {Max}, {Shields}, {Smith}
   \& {Ammons}}{{Rosario} et~al.}{2011}]{rosario11}
{Rosario} D.~J.,  {McGurk} R.~C.,  {Max} C.~E.,  {Shields} G.~A.,  {Smith}
  K.~L.,   {Ammons} S.~M.,  2011, \mn@doi [\apj] {10.1088/0004-637X/739/1/44},
  \href {https://ui.adsabs.harvard.edu/abs/2011ApJ...739...44R} {739, 44}

\bibitem[\protect\citeauthoryear{{Saglia} et~al.,}{{Saglia}
  et~al.}{2016}]{saglia16}
{Saglia} R.~P.,  et~al., 2016, \mn@doi [\apj] {10.3847/0004-637X/818/1/47},
  \href {https://ui.adsabs.harvard.edu/abs/2016ApJ...818...47S} {818, 47}

\bibitem[\protect\citeauthoryear{{Sahu}, {Graham}  \& {Davis}}{{Sahu}
  et~al.}{2019}]{sahu19}
{Sahu} N.,  {Graham} A.~W.,   {Davis} B.~L.,  2019, \mn@doi [\apj]
  {10.3847/1538-4357/ab0f32}, \href
  {https://ui.adsabs.harvard.edu/abs/2019ApJ...876..155S} {876, 155}

\bibitem[\protect\citeauthoryear{{Savorgnan} \& {Graham}}{{Savorgnan} \&
  {Graham}}{2016}]{savorgnan16}
{Savorgnan} G.~A.~D.,  {Graham} A.~W.,  2016, \mn@doi [\apjs]
  {10.3847/0067-0049/222/1/10}, \href
  {https://ui.adsabs.harvard.edu/abs/2016ApJS..222...10S} {222, 10}

\bibitem[\protect\citeauthoryear{{Schawinski}, {Treister}, {Urry}, {Cardamone},
  {Simmons}  \& {Yi}}{{Schawinski} et~al.}{2011}]{schawinski11}
{Schawinski} K.,  {Treister} E.,  {Urry} C.~M.,  {Cardamone} C.~N.,  {Simmons}
  B.,   {Yi} S.~K.,  2011, \mn@doi [\apjl] {10.1088/2041-8205/727/2/L31}, \href
  {https://ui.adsabs.harvard.edu/abs/2011ApJ...727L..31S} {727, L31}

\bibitem[\protect\citeauthoryear{{Schulze} et~al.,}{{Schulze}
  et~al.}{2015}]{schulze15}
{Schulze} A.,  et~al., 2015, \mn@doi [\mnras] {10.1093/mnras/stu2549}, \href
  {https://ui.adsabs.harvard.edu/abs/2015MNRAS.447.2085S} {447, 2085}

\bibitem[\protect\citeauthoryear{{Scoville} et~al.,}{{Scoville}
  et~al.}{2007}]{scoville07}
{Scoville} N.,  et~al., 2007, \mn@doi [\apjs] {10.1086/516585}, \href
  {http://adsabs.harvard.edu/abs/2007ApJS..172....1S} {172, 1}

\bibitem[\protect\citeauthoryear{{Shankar}, {Salucci}, {Granato}, {De Zotti}
  \& {Danese}}{{Shankar} et~al.}{2004}]{shankar04}
{Shankar} F.,  {Salucci} P.,  {Granato} G.~L.,  {De Zotti} G.,   {Danese} L.,
  2004, \mn@doi [\mnras] {10.1111/j.1365-2966.2004.08261.x}, \href
  {https://ui.adsabs.harvard.edu/abs/2004MNRAS.354.1020S} {354, 1020}

\bibitem[\protect\citeauthoryear{{Shankar}, {Weinberg}  \&
  {Miralda-Escud{\'e}}}{{Shankar} et~al.}{2009a}]{shankar09a}
{Shankar} F.,  {Weinberg} D.~H.,   {Miralda-Escud{\'e}} J.,  2009a, \mn@doi
  [\apj] {10.1088/0004-637X/690/1/20}, \href
  {http://adsabs.harvard.edu/abs/2009ApJ...690...20S} {690, 20}

\bibitem[\protect\citeauthoryear{{Shankar}, {Bernardi}  \& {Haiman}}{{Shankar}
  et~al.}{2009b}]{shankar09c}
{Shankar} F.,  {Bernardi} M.,   {Haiman} Z.,  2009b, \mn@doi [\apj]
  {10.1088/0004-637X/694/2/867}, \href
  {https://ui.adsabs.harvard.edu/abs/2009ApJ...694..867S} {694, 867}

\bibitem[\protect\citeauthoryear{{Shankar}, {Weinberg}  \& {Shen}}{{Shankar}
  et~al.}{2010a}]{shankar10b}
{Shankar} F.,  {Weinberg} D.~H.,   {Shen} Y.,  2010a, \mn@doi [\mnras]
  {10.1111/j.1365-2966.2010.16801.x}, \href
  {https://ui.adsabs.harvard.edu/abs/2010MNRAS.406.1959S} {406, 1959}

\bibitem[\protect\citeauthoryear{{Shankar}, {Crocce}, {Miralda-Escud{\'e}},
  {Fosalba}  \& {Weinberg}}{{Shankar} et~al.}{2010b}]{shankar10}
{Shankar} F.,  {Crocce} M.,  {Miralda-Escud{\'e}} J.,  {Fosalba} P.,
  {Weinberg} D.~H.,  2010b, \mn@doi [\apj] {10.1088/0004-637X/718/1/231}, \href
  {https://ui.adsabs.harvard.edu/abs/2010ApJ...718..231S} {718, 231}

\bibitem[\protect\citeauthoryear{{Shankar}, {Weinberg}  \&
  {Miralda-Escud{\'e}}}{{Shankar} et~al.}{2013}]{shankar13b}
{Shankar} F.,  {Weinberg} D.~H.,   {Miralda-Escud{\'e}} J.,  2013, \mn@doi
  [\mnras] {10.1093/mnras/sts026}, \href
  {https://ui.adsabs.harvard.edu/abs/2013MNRAS.428..421S} {428, 421}

\bibitem[\protect\citeauthoryear{{Shankar} et~al.,}{{Shankar}
  et~al.}{2016}]{shankar16}
{Shankar} F.,  et~al., 2016, \mn@doi [\mnras] {10.1093/mnras/stw678}, \href
  {http://adsabs.harvard.edu/abs/2016MNRAS.460.3119S} {460, 3119}

\bibitem[\protect\citeauthoryear{{Shankar} et~al.,}{{Shankar}
  et~al.}{2019}]{shankar19}
{Shankar} F.,  et~al., 2019, \mn@doi [\mnras] {10.1093/mnras/stz376}, \href
  {https://ui.adsabs.harvard.edu/abs/2019MNRAS.485.1278S} {485, 1278}

\bibitem[\protect\citeauthoryear{{Shankar} et~al.,}{{Shankar}
  et~al.}{2020}]{shankar20}
{Shankar} F.,  et~al., 2020, \mn@doi [Nature Astronomy]
  {10.1038/s41550-019-0949-y}, \href
  {https://ui.adsabs.harvard.edu/abs/2020NatAs...4..282S} {4, 282}

\bibitem[\protect\citeauthoryear{{Shen} et~al.,}{{Shen} et~al.}{2007}]{shen07}
{Shen} Y.,  et~al., 2007, \mn@doi [\aj] {10.1086/513517}, \href
  {https://ui.adsabs.harvard.edu/abs/2007AJ....133.2222S} {133, 2222}

\bibitem[\protect\citeauthoryear{{Shen} et~al.,}{{Shen} et~al.}{2013}]{shen13}
{Shen} Y.,  et~al., 2013, \mn@doi [\apj] {10.1088/0004-637X/778/2/98}, \href
  {http://adsabs.harvard.edu/abs/2013ApJ...778...98S} {778, 98}

\bibitem[\protect\citeauthoryear{{Shen} et~al.,}{{Shen} et~al.}{2015}]{shen15}
{Shen} Y.,  et~al., 2015, \mn@doi [\apj] {10.1088/0004-637X/805/2/96}, \href
  {https://ui.adsabs.harvard.edu/abs/2015ApJ...805...96S} {805, 96}

\bibitem[\protect\citeauthoryear{{Sheth}, {Mo}  \& {Tormen}}{{Sheth}
  et~al.}{2001}]{sheth01}
{Sheth} R.~K.,  {Mo} H.~J.,   {Tormen} G.,  2001, \mn@doi [\mnras]
  {10.1046/j.1365-8711.2001.04006.x}, \href
  {http://adsabs.harvard.edu/abs/2001MNRAS.323....1S} {323, 1}

\bibitem[\protect\citeauthoryear{{Silk} \& {Rees}}{{Silk} \&
  {Rees}}{1998}]{silk_rees98}
{Silk} J.,  {Rees} M.~J.,  1998, \aap, \href
  {http://adsabs.harvard.edu/abs/1998A%26A...331L...1S} {331, L1}

\bibitem[\protect\citeauthoryear{{Sinha} \& {Garrison}}{{Sinha} \&
  {Garrison}}{2020}]{sinha20}
{Sinha} M.,  {Garrison} L.~H.,  2020, \mn@doi [\mnras] {10.1093/mnras/stz3157},
  \href {https://ui.adsabs.harvard.edu/abs/2020MNRAS.491.3022S} {491, 3022}

\bibitem[\protect\citeauthoryear{{Suh}, {Civano}, {Trakhtenbrot}, {Shankar},
  {Hasinger}, {Sand ers}  \& {Allevato}}{{Suh} et~al.}{2020}]{suh20}
{Suh} H.,  {Civano} F.,  {Trakhtenbrot} B.,  {Shankar} F.,  {Hasinger} G.,
  {Sand ers} D.~B.,   {Allevato} V.,  2020, \mn@doi [\apj]
  {10.3847/1538-4357/ab5f5f}, \href
  {https://ui.adsabs.harvard.edu/abs/2020ApJ...889...32S} {889, 32}

\bibitem[\protect\citeauthoryear{{Terrazas}, {Bell}, {Henriques}, {White},
  {Cattaneo}  \& {Woo}}{{Terrazas} et~al.}{2016}]{terrazas16}
{Terrazas} B.~A.,  {Bell} E.~F.,  {Henriques} B. M.~B.,  {White} S. D.~M.,
  {Cattaneo} A.,   {Woo} J.,  2016, \mn@doi [\apjl]
  {10.3847/2041-8205/830/1/L12}, \href
  {https://ui.adsabs.harvard.edu/abs/2016ApJ...830L..12T} {830, L12}

\bibitem[\protect\citeauthoryear{{Ueda}, {Akiyama}, {Hasinger}, {Miyaji}  \&
  {Watson}}{{Ueda} et~al.}{2014}]{ueda14}
{Ueda} Y.,  {Akiyama} M.,  {Hasinger} G.,  {Miyaji} T.,   {Watson} M.~G.,
  2014, \mn@doi [\apj] {10.1088/0004-637X/786/2/104}, \href
  {https://ui.adsabs.harvard.edu/abs/2014ApJ...786..104U} {786, 104}

\bibitem[\protect\citeauthoryear{{Viitanen}, {Allevato}, {Finoguenov},
  {Bongiorno}, {Cappelluti}, {Gilli}, {Miyaji}  \& {Salvato}}{{Viitanen}
  et~al.}{2019}]{viitanen19}
{Viitanen} A.,  {Allevato} V.,  {Finoguenov} A.,  {Bongiorno} A.,  {Cappelluti}
  N.,  {Gilli} R.,  {Miyaji} T.,   {Salvato} M.,  2019, \mn@doi [\aap]
  {10.1051/0004-6361/201935186}, \href
  {https://ui.adsabs.harvard.edu/abs/2019A&A...629A..14V} {629, A14}

\bibitem[\protect\citeauthoryear{{White}, {Martini}  \& {Cohn}}{{White}
  et~al.}{2008}]{white08}
{White} M.,  {Martini} P.,   {Cohn} J.~D.,  2008, \mn@doi [\mnras]
  {10.1111/j.1365-2966.2008.13817.x}, \href
  {https://ui.adsabs.harvard.edu/abs/2008MNRAS.390.1179W} {390, 1179}

\bibitem[\protect\citeauthoryear{{Wyithe} \& {Loeb}}{{Wyithe} \&
  {Loeb}}{2009}]{wyithe09}
{Wyithe} J. S.~B.,  {Loeb} A.,  2009, \mn@doi [\mnras]
  {10.1111/j.1365-2966.2009.14647.x}, \href
  {https://ui.adsabs.harvard.edu/abs/2009MNRAS.395.1607W} {395, 1607}

\bibitem[\protect\citeauthoryear{{Zanisi} et~al.,}{{Zanisi}
  et~al.}{2021}]{zanisi21}
{Zanisi} L.,  et~al., 2021, \mn@doi [\mnras] {10.1093/mnras/stab1472}, \href
  {https://ui.adsabs.harvard.edu/abs/2021MNRAS.505.4555Z} {505, 4555}

\bibitem[\protect\citeauthoryear{{de Nicola}, {Marconi}  \& {Longo}}{{de
  Nicola} et~al.}{2019}]{denicola19}
{de Nicola} S.,  {Marconi} A.,   {Longo} G.,  2019, \mn@doi [\mnras]
  {10.1093/mnras/stz2472}, \href
  {https://ui.adsabs.harvard.edu/abs/2019MNRAS.490..600D} {490, 600}

\bibitem[\protect\citeauthoryear{{van den Bosch}}{{van den
  Bosch}}{2002}]{vandenbosch02}
{van den Bosch} F.~C.,  2002, \mn@doi [\mnras]
  {10.1046/j.1365-8711.2002.05171.x}, \href
  {http://adsabs.harvard.edu/abs/2002MNRAS.331...98V} {331, 98}

\bibitem[\protect\citeauthoryear{{van den Bosch}}{{van den
  Bosch}}{2016}]{vandenbosch16}
{van den Bosch} R. C.~E.,  2016, \mn@doi [\apj] {10.3847/0004-637X/831/2/134},
  \href {https://ui.adsabs.harvard.edu/abs/2016ApJ...831..134V} {831, 134}

\makeatother
\end{thebibliography}
\input{out/ms.bbl}
\bsp
\label{lastpage}

\end{document}